\documentclass[prd,preprintnumbers,pacs,12pt]{revtex4}
\usepackage{epsfig}
\usepackage{graphicx}
\usepackage {axodraw}

\usepackage{bm}
\usepackage{amsmath}
\usepackage{amssymb}
\usepackage{amscd}
\usepackage{latexsym}

\newcommand{\be}{\begin{equation}}

\newcommand{\ee}{\end{equation}}
\newcommand{\bea}{\begin{eqnarray}}
\newcommand{\eea}{\end{eqnarray}}

\newcommand{\nn}{\nonumber}
\newcommand{\de}{\partial}
\def\Black{}
 \def\AliasBlue{}
 \def\Blue{}
 \def\Brown{}

\begin{document}

\newcommand{\bra}[1]{\langle #1|}
\newcommand{\ket}[1]{|#1\rangle}
\newcommand{\braket}[2]{\langle #1|#2\rangle}
\newcommand{\tr}{\textrm{Tr}}
\newcommand{\lag}{\mathcal{L}}
\newcommand{\mbf}[1]{\mathbf{#1}}
\newcommand{\desl}{\slashed{\partial}}
\newcommand{\Desl}{\slashed{D}}

\renewcommand{\bottomfraction}{0.7}
\newcommand{\epsi}{\varepsilon}

\newcommand{\nl}{\nonumber \\}
\newcommand{\tc}[1]{\textcolor{#1}}
\newcommand{\sla}{\not \!}
\newcommand{\spinor}[1]{\left< #1 \right>}
\newcommand{\cspinor}[1]{\left< #1 \right>^*}
\newcommand{\Log}[1]{\log \left( #1\right) }
\newcommand{\Logq}[1]{\log^2 \left( #1\right) }
\newcommand{\mr}[1]{\mathrm{#1}}
\newcommand{\cw}{c_\mathrm{w}}
\newcommand{\sw}{s_\mathrm{w}}
\newcommand{\ct}{c_\theta}
\newcommand{\st}{s_\theta}
\newcommand{\gt}{{\tilde g}}
\newcommand{\gtp}{{{\tilde g}^\prime}}
\renewcommand{\i}{\mathrm{i}}
\renewcommand{\Re}{\mathrm{Re}}
\newcommand{\yText}[3]{\rText(#1,#2)[][l]{#3}}
\newcommand{\xText}[3]{\put(#1,#2){#3}}


\def\to{\rightarrow}
\def\ptl{\partial}
\def\beq{\begin{equation}}
\def\eeq{\end{equation}}
\def\bea{\begin{eqnarray}}
\def\eea{\end{eqnarray}}
\def\nn{\nonumber}
\def\half{{1\over 2}}
\def\rhalf{{1\over \sqrt 2}}
\def\calo{{\cal O}}
\def\call{{\cal L}}
\def\calm{{\cal M}}
\def\del{\delta}
\def\eps{\epsilon}
\def\lam{\lambda}

\def\anti{\overline}
\def\delfac{\sqrt{{2(\del-1)\over 3(\del+2)}}}
\def\heff{h'}
\def\square{\boxxit{0.4pt}{\fillboxx{7pt}{7pt}}\hspace*{1pt}}
    \def\boxxit#1#2{\vbox{\hrule height #1 \hbox {\vrule width #1
             \vbox{#2}\vrule width #1 }\hrule height #1 } }
    \def\fillboxx#1#2{\hbox to #1{\vbox to #2{\vfil}\hfil}   }

\def\braket#1#2{\langle #1| #2\rangle}
\def\gev{~{\rm GeV}}
\def\gam{\gamma}
\def\sn{s_{\vec n}}
\def\sm{s_{\vec m}}
\def\mm{m_{\vec m}}
\def\mn{m_{\vec n}}
\def\mh{m_h}
\def\sumn{\sum_{\vec n>0}}
\def\summ{\sum_{\vec m>0}}
\def\vl{\vec l}
\def\vk{\vec k}
\def\ml{m_{\vl}}
\def\mk{m_{\vk}}
\def\gp{g'}
\def\gt{\tilde g}
\def\hw{{\hat W}}
\def\hz{{\hat Z}}
\def\ha{{\hat A}}

\def\yy{{\cal Y}_\mu}
\def\yyt{{\tilde{\cal Y}}_\mu}
\def\lq{\left [}
\def\rq{\right ]}
\def\dmu{\partial_\mu}
\def\dnu{\partial_\nu}
\def\dmus{\partial^\mu}
\def\dnus{\partial^\nu}
\def\gp{g'}
\def\gpt{{\tilde g'}}
\def\gs{g''}
\def\ggs{\frac{g}{\gs}}
\def\eps{{\epsilon}}
\def\tr{{\rm {tr}}}
\def\V{{\bf{V}}}
\def\W{{\bf{W}}}
\def\Wt{\tilde{ {W}}}
\def\Y{{\bf{Y}}}
\def\Yt{\tilde{ {Y}}}
\def\L{{\cal L}}
\def\s{s_\theta}
\def\st{s_{\tilde\theta}}
\def\c{c_\theta}
\def\ct{c_{\tilde\theta}}
\def\gt{\tilde g}
\def\et{\tilde e}
\def\At{\tilde A}
\def\Zt{\tilde Z}
\def\Wpt{{\tilde W}^+}
\def\Wmt{{\tilde W}^-}

\newcommand{\Apt}{{\tilde A}_1^+}
\newcommand{\Bpt}{{\tilde A}_2^+}
\newcommand{\Amt}{{\tilde A}_1^-}
\newcommand{\Bmt}{{\tilde A}_2^-}
\newcommand{\Wtp}{{\tilde W}^+}
\newcommand{\Atp}{{\tilde A}_1^+}
\newcommand{\Btp}{{\tilde A}_2^+}
\newcommand{\Atm}{{\tilde A}_1^-}
\newcommand{\Btm}{{\tilde A}_2^-}
\def\mathswitchr#1{\relax\ifmmode{\mathrm{#1}}\else$\mathrm{#1}$\fi}
\newcommand{\Pe}{\mathswitchr e}
\newcommand{\Pp}{\mathswitchr {p}}
\newcommand{\PZ}{\mathswitchr Z}
\newcommand{\PW}{\mathswitchr W}
\newcommand{\PD}{\mathswitchr D}
\newcommand{\PU}{\mathswitchr U}
\newcommand{\PQ}{\mathswitchr Q}
\newcommand{\Pd}{\mathswitchr d}
\newcommand{\Pu}{\mathswitchr u}
\newcommand{\Ps}{\mathswitchr s}
\newcommand{\Pc}{\mathswitchr c}
\newcommand{\Pt}{\mathswitchr t}
\newcommand{\rd}{{\mathrm{d}}}
\newcommand{\GW}{\Gamma_{\PW}}
\newcommand{\GZ}{\Gamma_{\PZ}}
\newcommand{\GeV}{\unskip\,\mathrm{GeV}}
\newcommand{\MeV}{\unskip\,\mathrm{MeV}}
\newcommand{\TeV}{\unskip\,\mathrm{TeV}}
\newcommand{\fba}{\unskip\,\mathrm{fb}}
\newcommand{\pba}{\unskip\,\mathrm{pb}}
\newcommand{\nba}{\unskip\,\mathrm{nb}}
\newcommand{\PT}{P_{\mathrm{T}}}
\newcommand{\PTmiss}{P_{\mathrm{T}}^{\mathrm{miss}}}
\newcommand{\CM}{\mathrm{CM}}
\newcommand{\inv}{\mathrm{inv}}
\newcommand{\sig}{\mathrm{sig}}
\newcommand{\tot}{\mathrm{tot}}
\newcommand{\backg}{\mathrm{backg}}
\newcommand{\evt}{\mathrm{evt}}
\def\mathswitch#1{\relax\ifmmode#1\else$#1$\fi}
\newcommand{\M}{\mathswitch {M}}
\newcommand{\R}{\mathswitch {R}}
\newcommand{\TEV}{\mathswitch {TEV}}
\newcommand{\LHC}{\mathswitch {LHC}}
\newcommand{\MW}{\mathswitch {M_\PW}}
\newcommand{\MZ}{\mathswitch {M_\PZ}}
\newcommand{\Mt}{\mathswitch {M_\Pt}}
\def\si{\sigma}
\def\beqar{\begin{eqnarray}}
\def\eeqar{\end{eqnarray}}
\def\refeq#1{\mbox{(\ref{#1})}}
\def\reffi#1{\mbox{Fig.~\ref{#1}}}
\def\reffis#1{\mbox{Figs.~\ref{#1}}}
\def\refta#1{\mbox{Table~\ref{#1}}}
\def\reftas#1{\mbox{Tables~\ref{#1}}}
\def\refse#1{\mbox{Sect.~\ref{#1}}}
\def\refses#1{\mbox{Sects.~\ref{#1}}}
\def\refapps#1{\mbox{Apps.~\ref{#1}}}
\def\refapp#1{\mbox{App.~\ref{#1}}}
\def\citere#1{\mbox{Ref.~\cite{#1}}}
\def\citeres#1{\mbox{Refs.~\cite{#1}}}

\def\Black{}
 \def\AliasBlue{}
 \def\Blue{}
 \def\Brown{}

\title{Improved analysis of the bounds from the electroweak precision tests on the 4-site model}

\date{\today}
\author{Elena Accomando}%
 \email{E.Accomando@soton.ac.uk}
\affiliation{NExT Institute and School of Physics and Astronomy, University of Southampton, Highfield,
Southampton SO17 1BJ, UK}%
\author{Diego Becciolini}%
 \email{diego.becciolini@soton.ac.uk}
 \affiliation{NExT Institute and School of Physics and Astronomy, University of Southampton, Highfield,
Southampton SO17 1BJ, UK}%
 \author{Luca Fedeli}%
\affiliation{Universit\'a degli Studi di Firenze, Dip. di
Fisica e Astronomia, Firenze, Italy\\
and Istituto Nazionale di Fisica Nucleare, Sezione di Firenze, Italy}%
\author{Stefania De Curtis}%
 \email{decurtis@fi.infn.it}
\affiliation{Istituto Nazionale di Fisica Nucleare, Sezione di Firenze, Italy}%
\author{Daniele Dominici}%
\email{dominici@fi.infn.it, fedeli@fi.infn.it}
\affiliation{Universit\'a degli Studi di Firenze, Dip. di
Fisica e Astronomia, Firenze, Italy\\
and Istituto Nazionale di Fisica Nucleare, Sezione di Firenze, Italy}%

\begin{abstract}
\noindent We present a new complete analysis of the electroweak
precision observables within the recently proposed 4-site Higgsless
model, which is based on the $SU(2)_L\times SU(2)_1\times
SU(2)_2\times U(1)_Y$ gauge symmetry and predicts six extra gauge
bosons, $W^\pm_{1,2}$ and $Z_{1,2}$. Within the $\varepsilon_i$
(i=1,2,3,b) parametrization, we compute for the first time the EWPT
bounds via a complete numerical algorithm going beyond commonly used
approximations. Both $\varepsilon_{1,3}$ impose strong constraints.
Hence, it is mandatory to consider them jointly when extracting EWPT
bounds and to fully take in to account the correlations among the
electroweak precision measurements. The phenomenological consequence
is that the extra gauge bosons must be heavier than 250 GeV. Their
couplings to SM fermions, even if bounded, might be of the same
order of magnitude than the SM ones. In contrast to other Higgsless
models, the 4-site model is not fermiophobic. The new gauge bosons
could thus be discovered in the favoured Drell-Yan channel already
during the present run of the LHC experiment.
\end{abstract}

\maketitle

\section{Introduction}

In the past years a remarkable activity has been devoted to
investigate electroweak models formulated in extra dimension space
\cite{ArkaniHamed:1998rs,Antoniadis:1998ig,Randall:1999ee,
Csaki:2003dt,Agashe:2003zs,Csaki:2003zu,Barbieri:2003pr,Nomura:2003du,Cacciapaglia:2004jz,Cacciapaglia:2004rb,Contino:2006nn}.
In most of these scenarios, the size and shape of the extra
dimension(s) are responsible for solving the large hierarchy problem
and they can also provide viable alternatives to the Higgs
mechanism.  For example, in models where the standard model (SM)
gauge fields propagate in a fifth dimension, masses for the $W^\pm$
and $Z$ bosons can be generated via non-trivial boundary conditions
\cite{Csaki:2003dt,Cacciapaglia:2004jz,Cacciapaglia:2004rb,Nomura:2003du,Csaki:2003zu}.
Since the need for scalar doublets is eliminated in such scenarios,
these models have been aptly dubbed {\it Higgsless models}.  The
result of allowing the SM gauge fields to propagate in the bulk,
however, is towers of physical, massive vector gauge bosons (VGBs),
the lightest of which are identified with the SM $W^\pm$ and $Z$
bosons.  The heavier Kaluza-Klein (KK) modes, which have the
$SU(2)\times U(1)$ quantum numbers of the SM  $W^\pm$ and $Z$, play
an important role in longitudinal VGB scattering.  In the SM without
a Higgs boson, the scattering amplitudes for these processes
typically violate unitarity around $\sim$ 1 TeV \cite{Lee:1977eg}.
The exchange of light Higgs bosons, however, cancels the
unitarity-violating terms and ensures perturbativity of the theory
up to high scales.  In extra-dimensional Higgsless models, the
exchange of the heavier KK gauge bosons plays the role of the Higgs
boson and cancels the dominant  unitarity-violating terms
\cite{SekharChivukula:2001hz,Csaki:2003dt,Ohl:2003dp,Papucci:2004ip,Muck:2004br}.  As a result, the scale
of unitarity violation can be pushed upward in the TeV range.

The main drawback of extra-dimensional models is that they are
non-renormalizable and must be viewed as effective theories up
to some cut-off scale $\Lambda$ above which new physics must take over.
An extremely efficient and convenient way of studying the phenomenology
of five-dimensional effective theories in the context of
four-dimensional gauge theories is that of deconstruction. In fact
the discretization of the compact fifth dimension to a lattice generates the
so-called deconstructed theories which are chiral Lagrangian with a number of
replicas of the gauge group equal to the number of lattice sites
\cite{ArkaniHamed:2001ca,Arkani-Hamed:2001nc,Hill:2000mu,Cheng:2001vd,Abe:2002rj,Falkowski:2002cm,Randall:2002qr,Son:2003et,deBlas:2006fz}.
Models have been proposed, assuming a $SU(2)_L\times SU(2)_R\times U(1)_{B-L}$
gauge group in the 5D bulk,
\cite{Csaki:2003dt,Agashe:2003zs,Csaki:2003zu,Barbieri:2003pr,
Cacciapaglia:2004zv,Cacciapaglia:2004rb,Cacciapaglia:2004jz,Contino:2006nn},
in the framework suggested by the AdS/CFT correspondence, or also with a
simpler gauge group $SU(2)$ in the bulk
\cite{Foadi:2003xa,Hirn:2004ze,Casalbuoni:2004id,Chivukula:2004pk,Georgi:2004iy,Casalbuoni:2007xn}.
Deconstructed models possess extended gauge symmetries which
approximate the fifth dimension, but can be studied in the simplified
language of coupled non-linear $\sigma$-models
\cite{Appelquist:1980vg,Longhitano:1980iz,Longhitano:1980tm}.  In fact,
this method allows one to effectively separate the perturbativity calculable
contributions to low-energy observables from the strongly-coupled
contributions due to physics above $\Lambda$.  The former arise
from the new weakly-coupled gauge states, while the latter can be parameterized
by adding higher-dimensional operators \cite{Appelquist:1980vg,Longhitano:1980iz,Longhitano:1980tm,Bagger:1992vu,
Perelstein:2004sc,Chivukula:2007ic}.

The phenomenology of deconstructed Higgsless models has been
well-studied
\cite{Foadi:2003xa,Hirn:2004ze,Casalbuoni:2004id,Chivukula:2004pk,Perelstein:2004sc,
Georgi:2004iy,SekharChivukula:2004mu,SekharChivukula:2006cg}.  Recently, however, the
simplest version of these types of models, which involves only three
``sites'', has
received much attention and been shown to be capable of
approximating much of the interesting phenomenology associated with
extra-dimensional models and more complicated deconstructed
Higgsless models
\cite{Cacciapaglia:2004rb,Cacciapaglia:2005pa,Foadi:2004ps,Foadi:2005hz,
Chivukula:2005bn,Casalbuoni:2005rs,Chivukula:2005xm}.  The gauge
structure of the 3-site model is identical to that of the so-called
BESS (Breaking Electroweak Symmetry Strongly) which was first
analyzed more then twenty years ago
\cite{Casalbuoni:1985kq,Casalbuoni:1986vq}. Once electroweak
symmetry breaking (EWSB) occurs in the 3-site model, the gauge
sector consists of a massless photon, three relatively light massive
VGBs which are identified with the SM $W^\pm$ and $Z$ gauge bosons, as
well as three new heavy VGBs which we denote as $W_1^\pm$ and
$Z_1$. The exchange of these heavier states in longitudinal VGB
scattering can delay unitarity violation up to higher scales
(for discussions  of unitarization through new vector states, see
\cite{Bagger:1993zf,Casalbuoni:1997bv,Foadi:2003xa,Barbieri:2008cc,Barbieri:2009tx,Accomando:2008jh}).

The drawback of all these models, as with technicolor theories, is
to reconcile the presence of a relatively low KK-spectrum, necessary
to delay the unitarity violation to TeV-energies, with the
electroweak precision tests (EWPT) whose measurements can be
expressed as functions of the $\eps_1,\eps_2$ and $\eps_3$ (or $T, U,
S$) parameters
\cite{Peskin:1990zt,Peskin:1992sw,Altarelli:1991zd,Altarelli:1998et}. These parameters are defined in terms of the SM
gauge boson self-energies, $\Pi^{\mu\nu}_{ij}(q^2)$, where $(ij)$ =
$(WW), (ZZ), (\gamma\gamma)$ and $(Z\gamma)$, and $q$ is the
momentum carried by the external gauge bosons. More in detail, while
$\eps_1$ and $\eps_2$ are protected by the custodial symmetry,
shared by both the aforementioned classes of models, the $\eps_3$
($S$) parameter constitutes the real obstacle to EWPT consistency.
This problem can be solved by either delocalizing fermions along the
fifth dimension \cite{Cacciapaglia:2004rb,Foadi:2004ps} or,
equivalently in the deconstructed version of the model, by allowing
for direct couplings between new vector bosons and SM fermions
\cite{Casalbuoni:2005rs}. In the simplest version of this latter
class of models, corresponding to just three lattice sites and gauge
symmetry $SU(2)_L\times SU(2)\times U(1)_Y$ (the BESS model), the
requirement of vanishing of the $\eps_3$ parameter implies that the
new triplet of vector bosons is almost fermiophobic. As a
consequence, the only production channels where the new gauge bosons
can be searched for are those driven by boson-boson couplings. The
Higgsless literature has been thus mostly focused on difficult
multi-particle processes which require high luminosity to be
detected, that is vector boson fusion and associated production of
new gauge bosons with SM ones
\cite{Birkedal:2004au,Belyaev:2007ss,He:2007ge,Ohl:2008ri}.

The minimal 3-site model can be extended by inserting an additional lattice
site. The newly obtained next-to-minimal (4-site) Higgsless model is based on the
$SU(2)_L\times SU(2)_1\times SU(2)_2\times U(1)_Y$ gauge symmetry. It predicts
two neutral and four charged extra gauge bosons, $Z_{1,2}$ and $W^\pm_{1,2}$,
and satisfies the EWPT constraints without necessarily having fermiophobic
resonances \cite{Accomando:2008jh,Accomando:2008dm,Accomando:2010ir}.
Within this framework, the more promising Drell-Yan processes become
particularly relevant for the extra gauge boson search at the TEVATRON and the LHC.

In this paper, we present a new calculation of the EWPT bounds on
the 4-site Higgsless model. There are two new ingredients compared
to the existing results present in the literature. The first one
concerns the computation of the 4-site Higgsless model contributions
to the $\varepsilon_i$ (i=1,2,3,b) parameters, which summarize the
electroweak precision measurements performed by LEP, SLD and
TEVATRON experiments. These contributions are computed for the first
time via a complete numerical algorithm, going beyond commonly used
analytical approximations. The second ingredient addresses the
minimum $\chi^2$ test, used to extract bounds on the 4-site model.
We improve previous simplified analysis, by taking into account the
full correlation between the measurements of all four
$\varepsilon_i$ (i=1,2,3) and $\varepsilon_b$
\cite{Altarelli:1993sz} parameters.
The effect of the correlations was already considered but within the 3-site model \cite{Abe:2008hb}.
 We moreover analyze the cutoff
dependence of EWPT bounds, and discuss how well the 4-site Higgsless
model can reproduce experimental results. We finally show the
portion of the parameter space which survives the EWPT. Within that
framework, we give a description of the main properties of the
additional four charged and two neutral gauge bosons predicted by
the 4-site Higgsless model.

The paper is structured as follows. In Sect.\ref{4-site}, we review the
next-to-minimal 4-site Higgsless model. In Sects.\ref{EWPT_new}-\ref{miai_s},
we update the bounds from the EWPT and we derive the new allowed
parameter space. Here, we define mass spectrum and gauge couplings
of the extra $Z_{1,2}$ and $W^\pm_{1,2}$ vector bosons.
In Sect.\ref{App_vs_Exac}, we compare the new exact results with
those obtained via common approximations. Finally in
Sect.\ref{3-site_s}, for completeness, we compute the new exact EWPT
bounds on the minimal 3-site Higgsless model, and we compare
quantitatively minimal and next-to-minimal Higgsless scenarios.
Conclusions are given in Sect.\ref{conclusions}.
Appendix A contains, as a reference, the approximate calculations.

\section{Review of the 4-site Higgsless model}
\label{4-site}

The class of models we are interested in follows the idea of dimensional deconstruction
\cite{ArkaniHamed:2001ca,Arkani-Hamed:2001nc,Hill:2000mu,Cheng:2001vd}
and was recently studied in \cite{Casalbuoni:2005rs}. The so-classified theories can also be seen as a generalization of the BESS model
\cite{Casalbuoni:1985kq,Casalbuoni:1986vq,Casalbuoni:1989xm} to an
arbitrary number of new triplets of gauge bosons. In their general
formulation
\cite{Foadi:2003xa,Hirn:2004ze,Casalbuoni:2004id,Chivukula:2004pk,Georgi:2004iy},
they are based on the  $SU(2)_L\otimes SU(2)^K\otimes U(1)_Y$ gauge
symmetry, and contain $K+1$ non-linear $\sigma$-model scalar fields
which trigger the spontaneous symmetry breaking.

The 4-site Higgsless model, described in Refs.\cite{Accomando:2010ir,Accomando:2008jh,Accomando:2008dm}, is defined by taking
$K$=2 and requiring the Left-Right (LR) symmetry in the gauge sector. In the unitary gauge, it predicts two new triplets of gauge bosons which
 acquire mass through the same symmetry breaking mechanism which gives mass to the SM gauge bosons. By calling
$\tilde{W}_{i\mu}=\tilde{W}_{i\mu}^{a}\tau^a/2$ and $g_i$ the gauge fields and couplings associated to the extra $SU(2)_i$, $i=1,2$;
$\Wt_\mu=\Wt_\mu^{a}\tau^a/2$, $\Yt_\mu=\yyt\tau^3/2$ and $\gt$,
$\gpt$ the gauge fields and couplings  associated to $SU(2)_L$ and
$U(1)_Y$ respectively, the charged gauge boson mass Lagrangian
is given by:
\begin{equation}\label{LMC}
{\mathcal L}^{{\mathcal C}}_{mass} = \tilde{{\mathcal C}}^-_\mu\,
{\mathcal M}^2_c\, \tilde{{\mathcal C}}^{\mu+}
\end{equation}
with $\tilde{{\mathcal C}^-}\!=\left(\!
\begin{array}{ccc}
\Wmt \!\!, &\!\! \tilde W_1^-\!\!, &\! \!\tilde W_2^-\\
\end{array}
\!\right) $, $\tilde{{\mathcal C}^+}=(\tilde{{\mathcal
C}^-})^\dagger$, and
 \be \label{M2}
 {\mathcal M}^2_c=\left(
\begin{array}{ccc}
{\gt^2\over g_1^2} M_1^2& - {\gt\over g_1} M_1^2& 0 \\
-{\gt\over g_1} M_1^2 & {1\over 2}(M_1^2 + M_2^2)  & {1\over 2}(M_1^2-M_2^2)\\
0 & {1\over 2}(M_1^2-M_2^2) & {1\over 2}(M_1^2 + M_2^2) \\
\end{array}
\right) \ee where $M_{1,2}$ are the bare masses of the six
additional gauge bosons, $W^\pm_{1,2}$, $Z_{1,2}$ and we had taken
$g_1=g_2$ in virtue of the LR symmetry imposed in the gauge sector.

Similarly, the mass Lagrangian of the neutral gauge sector is:
\begin{equation}\label{Lmn}
{\lag}^{\mathcal N}_{mass} = \frac{1}{2}\tilde N^T_\mu {\mathcal
M}^2_n \tilde{N}^\mu
\end{equation}
with $\tilde{N}^T = \left( \tilde{W^3}, \tilde{W_1^3},
\tilde{W_2^3}, \tilde{\mathcal{Y}} \right )$ and
\begin{equation}
 {\mathcal M}_n^2=\left(
\begin{array}{cccc}
{\gt^2\over g^2_1} M_1^2 & -{\gt\over g_1}M_1^2 & 0 & 0\\
-{\gt\over g_1}M_1^2 & {1\over 2}(M_1^2+M_2^2) & {1\over 2}(M_1^2-M_2^2) & 0\\
0 & {1\over 2}(M_1^2-M_2^2) & {1\over 2}(M_1^2+M_2^2) & -{\gpt\over g_1}M_1^2\\
0 & 0 & -{\gpt\over g_1}M_1^2 & {\tilde g'^{2}\over g_1^2}M_1^2\\

\end{array}
\right)
\end{equation}
Direct couplings of the new gauge bosons to SM fermions can be included
in a way that preserves the symmetry of the model. The fermion
Lagrangian is given by:

\bea {\cal L}_{fermions}&=&\bar\psi_L i\gamma^\mu \de_\mu \psi_L +
\bar\psi_R i\gamma^\mu \de_\mu \psi_R\nn\\
&-&\frac 1 {{1+b_1+b_2}}\bar\psi_L \gamma^\mu
 \gt\Wt_\mu\psi_L\nn\\
&-& \sum_{i=1}^2 \frac {b_i}{{1+b_1+b_2}}\bar\psi_L \gamma^\mu
 g_i \tilde{W}_{i\mu}\psi_L\nn\\
&-&\bar\psi_R\gamma^\mu  (\gpt \Yt_\mu+ \frac 1 2 \gpt(B-L)
\yyt)\psi_R-\bar\psi_L \gamma^\mu\frac 1 2 \gpt (B-L)
\yyt\psi_L.\label{eq:13}
\eea
In the above formula, $b_{1,2}$ are two arbitrary dimensionless parameters, which we assume to be the same for quarks and fermions of each
 generation, and $\psi_{L(R)}$ denotes the standard quarks and leptons. Direct couplings of the new gauge bosons to SM right-handed fermions
 could also be introduced. They are however strongly constrained by data from non-leptonic K-decays and $b\to s\gamma$ processes
 \cite{Yao:2006px} to be of order of $10^{-3}$ \cite{Bechi:2006sj}. For this reason, we neglect them.

The 4-site Higgsless model contains seven parameters a priori:
$\tilde g,\tilde g', g_1, M_1, M_2, b_1, b_2$. However, their number can be
reduced to four, by fixing the gauge couplings $\tilde g,\tilde g', g_1$ in
terms of the three SM input parameters $e, G_F, M_Z$ which denote
electric charge, Fermi constant and Z-boson mass, respectively.
As a result, our parameter space is defined by
four free parameters: $M_{1,2}$ which represent the bare masses of the lighter
($W_1^\pm, Z_1$) and heavier ($W_2^\pm, Z_2$) gauge boson triplets, and
$b_{1,2}$ which are their bare direct couplings to SM fermions. In the
following, we will give our results also in terms of $z=M_1/M_2$, the ratio
of the bare masses.

\subsection{Free parameters versus physical observables}

\begin{figure}[!htbp]
\begin{center}
\unitlength1.0cm
\begin{picture}(7,4)
\put(-5.6,-4.8){\epsfig{file=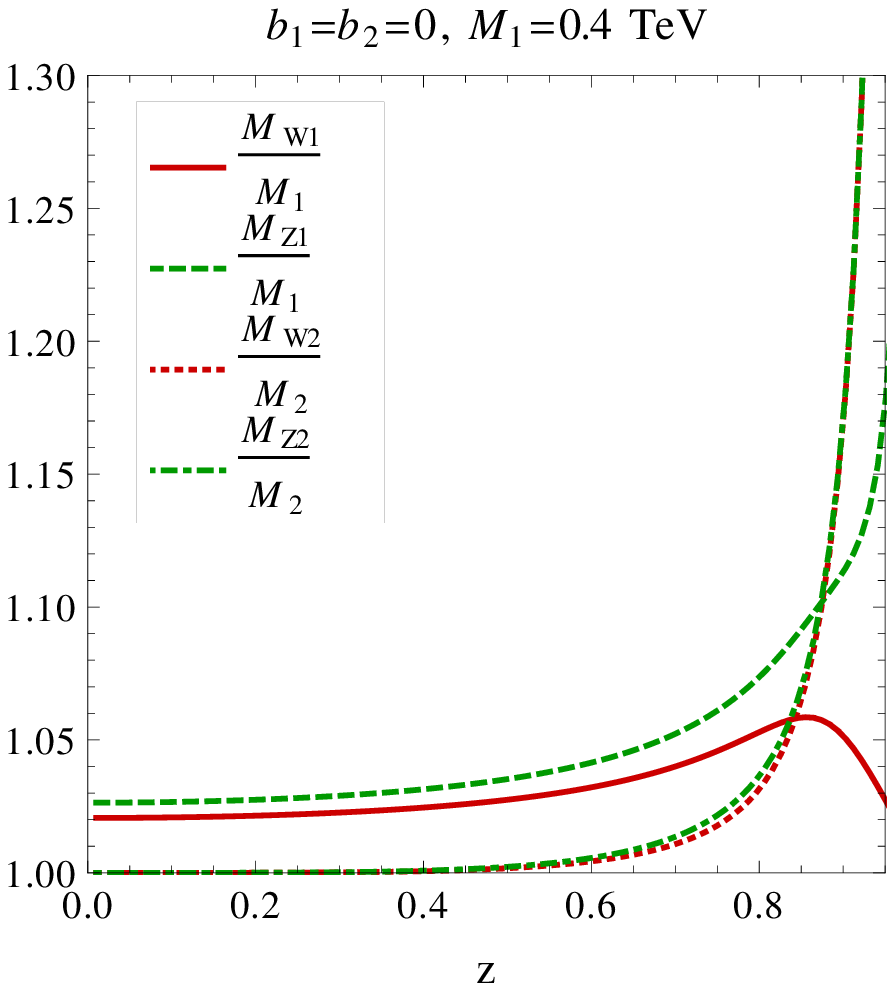,width=7.5cm}}
\put(3.5,-4.8){\epsfig{file=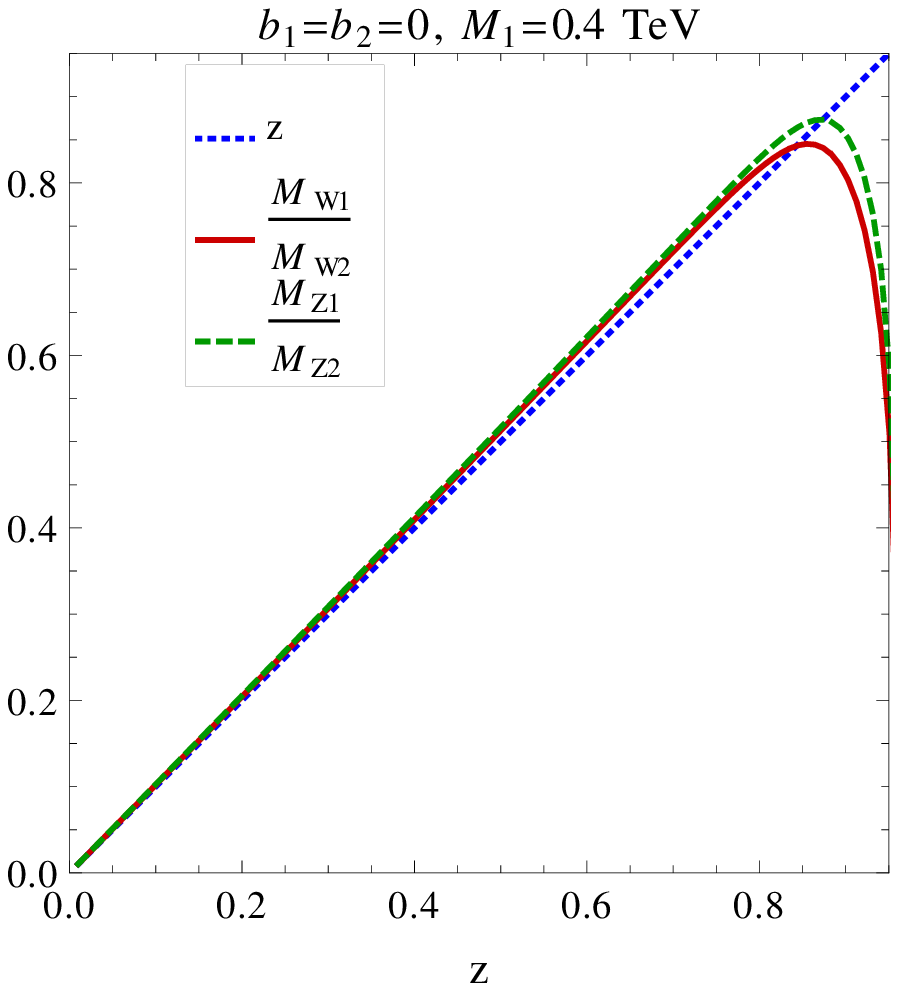,width=7.3cm}}
\end{picture}
\end{center}
\vskip 4.cm
\caption{Left: Ratios $M_{Wi}/M_{i}$, $M_{Zi}/M_{i}$ ($i=1,2$) as a function of $z=M_1/M_2$ at fixed $M_1$=0.4 TeV.
Right: Ratios $M_{W1}/M_{W2}$, $M_{Z1}/M_{Z2}$, and $z=M_1/M_2$ as a function of $z$ at fixed $M_1$=0.4 TeV. We fix $b_{1,2}=0$.}
\label{fig_z_Mi}
\end{figure}

Before starting the new analysis of the EWPT bounds on the 4-site
Higgsless model, it is useful to understand how the free parameters
of the model are connected to the physical quantities. We focus here
on the gauge sector (the fermionic one will be discussed later in
Sect.\ref{miai_s}) and we analyze the relation between mass
eigenvalues and bare masses, $M_{1,2}$. The results are displayed in
Fig.~\ref{fig_z_Mi}. In the left plot, we show the ratio between
physical and bare masses, $M_{Vi}/M_i$ ($V=W,Z$ and i=1,2), as a
function of $z=M_1/M_2$ for a given representative
 value $M_1$=0.4 TeV. Let us notice
that the mass eigenvalues acquire a dependence on the direct couplings between extra gauge bosons and ordinary matter, $b_{1,2}$, via the
 $G_F$ constraint. This dependence is however
quite mild. Thus, at fixed $M_1$, everything is driven by the $z$
parameter and we can safely fix $b_{1,2}=0$. From the left plot in
Fig.~\ref{fig_z_Mi}, one can see that the corrections to the bare
mass parameters are positive. More in detail, they do not exceed
$\mathcal O(5\%)$ for low-intermediate values of $z$, while they
sensibly increase up to $\mathcal O(30\%)$ for high $z$ values. This
behavior characterizes the low-intermediate mass spectrum, which the
chosen $M_1$=0.4 TeV value in Fig.~\ref{fig_z_Mi} is an example of.
The situation changes drastically, and gets more stable, if one
moves to larger mass scales. For $M_1\ge$ 1 TeV indeed the
corrections to the bare masses never exceed $O(5\%)$ over the full
$z$ range. We can observe a similar behavior in the ratio between
the masses of lighter and heavier extra gauge bosons. In the right
plot of
 Fig.~\ref{fig_z_Mi}, we compare $z=M_1/M_2$ with the corresponding ratios between the mass eigenvalues of charged and neutral extra gauge
 bosons.
We fix, as before, $M_1$=0.4 TeV and plot the three different ratios $z$, $M_{W1}/M_{W2}$ and $M_{Z1}/M_{Z2}$ versus $z$. Once again, the
 bare parameter $z=M_1/M_2$ appears to be a good approximation of the ratio between $M_{W1,Z1}$ and $M_{W2,Z2}$ except for low masses
$M_1\le$ 1 TeV and high $z$ values where it can overestimate the physical ratios up to about 40\%. In this latter region in fact the
corrections to $M_2$ are much stronger than those to $M_1$,
as shown in the left plot, giving rise to a sharp decrease in the $M_{W1}/M_{W2}$ and $M_{Z1}/M_{Z2}$ ratios compared to the bare $z$ value.
Thus summarizing, in the low-intermediate $z$ region, the bare parameters give an excellent description of the physical quantities,
accurate at percent level for low masses, and at permil level for
$\mathcal O$(TeV) masses. In the high $z$ region instead, the bare parameters
give a good estimate of the physical masses only for $M_1\ge$ 1 TeV, while
the low edge of the spectrum is poorly reproduced.

The above mentioned physical masses and couplings of the extra gauge bosons
to ordinary matter are obtained via a complete numerical
algorithm in terms of the four free parameters of the model: $M_1$, $z$, $b_1$, $b_2$. This represents a novelty compared to previous
publications \cite{Accomando:2010ir,Accomando:2008jh,Accomando:2008dm}. The outcome is the ability to reliably and accurately describe
the full parameter space of the 4-site Higgsless model even in regions of low mass and high $z$ where previously used approximations would
fail, as we will discuss in detail in Sect.\ref{App_vs_Exac}.

\section{Bounds from EWPT: update of the $\varepsilon_{1,2,3,b}$ analysis}
\label{EWPT_new}

Universal electroweak radiative corrections to the precision observables
measured by LEP, SLD and TEVATRON experiments can be efficiently quantified in terms
of three parameters: $\varepsilon_1, \varepsilon_2$, and $\varepsilon_3$ (or S,
T, and U) \cite{Peskin:1990zt,Peskin:1992sw,Altarelli:1991zd,Altarelli:1998et}.
 A fourth parameter, $\varepsilon_b$, can be added to describe non
universal effects associated to the bottom quark sector \cite{Altarelli:1993sz}.
Besides the SM contributions, also potential heavy new physics may affect the
low-energy electroweak precision data through these four parameters. For that
reason, the $\varepsilon_i$ (i=1,2,3,b) are a powerful method to constrain
theories beyond the SM. We use this parametrization to derive bounds on the
3-site (or BESS) and 4-site Higgsless models.
Measurements by the LEP2 experiment can be summarized in four additional
parameters $V,X,Y,W$ \cite{Barbieri:2004qk}. However, the 3-site and 4-site model
contributions to these observables are strongly suppressed. We thus neglect them,
and focus only on the $\varepsilon_i$ (i=1,2,3,b) parameters.

In the literature on Higgsless models, major attention has been devoted to the
$\varepsilon_3$ (or S) parameter. The computations have been performed mainly
at tree level, by making use of different approximations. The common feature
of these approximate results is that they all rely on a series expansion in
the ratio $e/g_1$, where $e$ is the electric charge and $g_1$ the extra gauge group coupling constant, and in the model parameters which
measure the amount of fermion delocalization in the five dimensional theory interpretation (in the deconstructed version they are represented
by the $b_i$ parameters).
In this approximation, the $\varepsilon_{1,2}$ parameters vanish at tree
level owing to the custodial symmetry, at least at the first order in the
fermion delocalization parameter expansion. This is the reason why most of
the physics community has focused on $\varepsilon_3$. In addition to the
discussed approximate tree level results, in the recent years preliminary
calculations of one-loop corrections have been performed. More in detail,
the one-loop chiral logarithmic corrections to the $\varepsilon_{1,3}$ (or T
and S) parameters have been evaluated for the 3-site and 4-site models
\cite{Matsuzaki:2006wn,SekharChivukula:2007ic,Dawson:2007yk,Dawson:2008as}.
At the present status of the $\varepsilon_i$ calculation, the
one-loop contribution to the $\varepsilon_1$ parameter is of course dominant.

In this paper, we aim to fill the gap between approximate tree level
results and attempts of improved precision at one-loop. We
concentrate on the tree level calculation, going beyond the popular
approximations summarized above. We thus compute the four
$\varepsilon_i$ (i=1,2,3,b) exactly, keeping their full dependence
on the model parameters, via a numerical algorithm. In order to
understand quantitatively the difference between exact and
approximate results, and maintain a link with the previous
literature, in Sec.\ref{App_vs_Exac} we will compare our exact
numerical calculation with the approximate expansion up to the
second order in the $e/g_1$ parameter, keeping the $b_i$ direct
coupling content exact. The physical motivation to go beyond the
first order perturbative expansion of the $\varepsilon_i$
(i=1,2,3,b) in the model parameters is three-fold. The first reason
is to give a complete description of the parameter space. As the
bare mass parameter $M_1$ is roughly proportional to the gauge
coupling $g_1$, and strictly linked to the physical masses
$M_{W1,Z1}$, in order to reach the low edge of the spectrum one has
to deal with small $g_1$ values where the expansion in $e/g_1$ is
not reliable anymore. In addition, the contributions to the
$\varepsilon_i$ coming from the direct couplings between SM fermions
and new vector bosons, $b_{1,2}$, either induced by the presence of
new heavy fermions or by the fermion delocalization in the bulk when
considering theories in five dimensions, can undergo delicate
cancelations. While in the 3-site model there is only one bare
direct coupling, and fine-tuned to keep the fermion couplings of the
new gauge bosons very small in order to accommodate EWPT (almost
fermiophobic scenario), in the 4-site extension of Higgsless models
there are two bare direct couplings, thus some interplay between
them, allowing for larger couplings within the bounds. In this
latter case, subtle cancelations take place and the perturbative
expansion up to the first order in the fermion-boson direct
couplings (or fermion delocalization parameter) is not good anymore.
Finally, $\varepsilon_{1,2,3}$ receive logarithmic loop corrections
from SM particles which increase with energy. Within the SM, such a
bad high energy behavior is cutoff by the mass of a light Higgs.
But, obviously, in Higgsless models these contributions become
extremely important when approaching $O$(TeV) energy scales. It is
thus necessary to compute precisely not only $\varepsilon_{3}$ but
also $\varepsilon_{1,2}$ in order to see whether the new physics,
alternative to the light elementary Higgs, can balance the bad SM
logarithmic growth with energy. For all these reasons, in order to
derive realistic and reliable bounds on Higgsless models it is
mandatory to exactly compute all $\varepsilon_i$ (i=1,2,3,b)
parameters, and perform a combined fit to the experimental results
taking into account their full correlation.

Triple gauge boson vertex bounds could give a lower limit 
on the masses of the heavier resonances as studied within the 3-site model 
\cite{SekharChivukula:2006cg} for ideal localization of  fermions. 
However in our model we have a modification not only of the trilinear 
$ZW^+W^-$ vertex but also of the couplings of $Z,W$ to fermions. Therefore 
LEP2 measurements on cross sections $e^+e^- \to W^+W^- \to $ 4 fermions can be 
used to obtain bounds on the 4-site parameter space but this requires a 
complete calculation of the cross section taking into account all 
these modifications and in principle also the exchange of the new resonances. 
All these effects 
have to be taken into account for a reliable
analysis of LEP2 bounds. This is beyond the scope of the present paper and 
we plan to pursue in a future publication.

\subsection{Computing $\varepsilon_1, \varepsilon_2, \varepsilon_3$, and
$\varepsilon_b$ in the 4-Site Higgsless model.}

The three electroweak $\varepsilon_i$ (i=1,2,3) parameters, summarizing the
universal electroweak corrections to the precision observables measured by
LEP, SLD and TEVATRON, can be obtained from $\Delta r_W$, $\Delta\rho$ and $\Delta k$
\cite{Altarelli:1993sz,Altarelli:1998et}:
\bea
\eps_1&=& \Delta\rho\,\nn\\
\eps_2 &=& \c^2\Delta\rho+\frac{\s^2}{c_{2\theta}}\Delta r_W-2 \s^2\Delta k\,\nn\\
\eps_3 &=& \c^2\Delta\rho+c_{2\theta}\Delta k \,
\label{epsdef}
\eea
with the Weinberg angle defined by
\be\label{Sin2} \s^2\c^2=\frac{{\sqrt{2}}
e^2}{8 M_Z^2G_F}.
\ee
In this scheme, the physical inputs are chosen to be the electric charge, the
Fermi constant and the $Z$-boson mass:
\bea
\sqrt{4\pi\alpha}&=& 0.3123\\
G_F&=& 1.16639\times 10^{-5} \, \mbox{GeV}^{-2}\\
M_Z &=& 91.1876 \, \mbox{GeV} \, \label{sminputs} \eea The Weinberg
angle is thus uniquely determined. The fourth $\varepsilon_b$
parameter, describing instead non-universal effects in the bottom
quark sector, is related to the corrections to the SM $Z$-boson
coupling to left-handed $b$-quarks, $\delta g_{Lb}$, as follows: \be
\varepsilon_b=-2\delta g_{Lb}. \ee Within the 4-site Higgsless
model, the $\varepsilon_b$ parameter is zero owing to family
universality in the fermionic sector. It receives however a
contribution from SM radiative corrections, and it is experimentally
correlated to the other three $\varepsilon_i$ (i=1,2,3) parameters.
For this reason, we analyze its effect. In principle a non
universality of direct couplings could be considered for the $(t,b)$
sector to describe a special role of this doublet due to its
possible compositeness
\cite{Miransky:1988xi,Miransky:1989ds,Bardeen:1989ds,Hill:1994hp,Hill:1991at,Han:2003pu,Agashe:2003zs,Contino:2006nn}.
In this paper we don't consider such an alternative.

In order to compute the new physics contributions to the $\varepsilon_i$
(i=1,2,3) parameters, we follow the procedure of diagonalizing the charged
and neutral mass matrices. We thus derive the mass eigenstates of the gauge
sector, and recast the Lagrangian in terms of those eigenvectors.
Once the Lagrangian given in Eq. (\ref{eq:13}) has been re-expressed in terms
of charged and
neutral gauge boson mass eigenstates, the two $\Delta\rho$ and $\Delta k$
parameters can be extracted from the neutral current couplings to the SM
$Z$-boson:
\be
\L^{neutral}(Z)=-\frac{e}{\s\c}
\Big(1+\frac{\Delta\rho}{2}\Big)Z_\mu\overline\psi [\gamma^\mu
g_V+\gamma^\mu \gamma_5g_A]\psi
\ee
with
\be
g_V =
\frac{\mathbf{T^3}}{2}-s^2_{\theta_{eff}} \mathbf{Q},\quad g_A =
-\frac{\mathbf{T^3}}{2},\quad s^2_{\theta_{eff}} = (1+\Delta k) \s^2.
\ee
The $\Delta r_W$ parameter is instead given by:
\be
\frac{M^2_W}{M^2_Z}=
 \c^2\left[1-\frac{\s^2}{c_{2\theta}}\Delta r_W\right]\,
\ee
where $M_W$ and $M_Z$ are the SM $W^\pm$ and $Z$ boson masses.

The tree level contribution of the 4-site Higgsless model to the
$\varepsilon_i$ (i=1,2,3) parameters has been computed exactly, via a
complete numerical calculation. This represents a novelty. In the literature,
in fact, these tree level new physics effects are evaluated via an analytical
truncated multiple expansion in the extra gauge coupling, $e/g_1$, and the direct
couplings of the extra gauge bosons with SM fermions (or delocalization
parameters), that is $b_{1,2}$ in our notation. The exact result we present
in this paper allows one to span the full parameter space of the model,
reliably computing also regions characterized by small $g_1$ (or $M_1$)
values, and sizable $b_{1,2}$ couplings where the common approximated
expansion would fail.
For sake of comparison, in Appendix A we derive the $\varepsilon_i$ parameters
via an analytical expansion up to the order $\mathcal O(e^2/g_1^2)$, keeping the
full $b_{1,2}$ content. In Sec.\ref{App_vs_Exac}, we discuss the goodness of this approximation, and define its validity domain by comparing
 it to the exact numerical solution.

\subsection{Fit to the ElectroWeak Precision Tests}

By making use of the electroweak precision observables measured by LEP, SLD and TEVATRON
experiments, one can determine the $\varepsilon_i$ (i=1,2,3,b) parameters
as \cite{:2005ema}
\begin{equation}\label{eq:LEP1}
\begin{array}{l}
\varepsilon_1^{exp}= +(5.4\pm 1.0)~10^{-3}\\
\varepsilon_{2}^{exp} = -(8.9\pm 1.2)~10^{-3}\\
\varepsilon_{3}^{exp} =+(5.34\pm 0.94)~10^{-3}\\
\varepsilon_{b}^{exp} =-(5.0\pm 1.6)~10^{-3}\\
\end{array}\,\,\,\,\,\,
\rho = \begin{pmatrix}1 & 0.60& 0.86 & 0.00\\ 0.60 & 1 & 0.40 & -0.01\\
0.86 & 0.40 & 1 & 0.02\\ 0.00 & -0.01 & 0.02 & 1
\end{pmatrix}
\end{equation}
where $\rho$ is the correlation matrix.
In order to perform a complete EWPT analysis and pose constraints on the
parameters of the 4-site Higgsless model, we need to include also the
SM universal electroweak radiative corrections to the four $\varepsilon_i$
parameters.
We make use of the following expressions, obtained
with the code TopaZ0  to compute the radiative corrections with $m_t^{pole}=172.7$ GeV \cite{Agashe:2005dk}:
\begin{eqnarray}
\varepsilon_1^{rad}&=&(+5.6-0.86 \ln\frac{M_H}{M_Z})10^{-3}\\
\varepsilon_2^{rad}&=&(-7.09+0.16 \ln\frac{M_H}{M_Z})10^{-3}\\
\varepsilon_3^{rad} &=& (+5.25+0.54 \ln\frac{M_H}{M_Z})10^{-3}\\
\varepsilon_b^{rad} &=& -6.43\, 10^{-3}.
\label{eps123l}
\end{eqnarray}
These equations represent an effective and sufficiently accurate numerical
approximation of the pure SM contribution. The Higgs mass, $M_H$, should be
interpreted in our model as an ultraviolet cutoff of the SM loops provided by
the model itself. These terms correspond to UV logarithms in the
low energy Higgsless theory. We will take $M_H=1,~3$ TeV.
The first case corresponds to the extrapolated SM predictions in presence
of a scalar bound state which saturates the Lee-Quigg-Thacker bound \cite{Lee:1977eg}.
The second corresponds to the case with no bound state and $M_H$ is
interpreted as the cutoff of the theory. For comparison with the SM fit, we
will consider also a case with $M_H=300$ GeV.

We are now ready to extract bounds on the free parameters of the 4-site
Higgsless model, $M_{1,2}$ and $b_{1,2}$, by performing a minimum $\chi^2$
test. The $\chi^2$ function is defined as:
$$
\chi^2 =\sum_{i,j} (\varepsilon_i+\varepsilon_i^{rad}-\varepsilon_i^{exp})
\left [(\sigma^2)^{-1}\right ]_{ij}(\varepsilon_j+\varepsilon_j^{rad}-\varepsilon_j^{exp}),
\qquad\hbox{where}
\qquad (\sigma^2)_{ij} = \sigma_i \rho_{ij} \sigma_j\ .$$
In the above equation, $\sigma_i$ is the standard deviation and $\rho_{ij}$
the correlation matrix of Eq.~\refeq{eq:LEP1}.
The global minimum $\chi^2$, obtained by minimizing with respect to the four
free parameters $M_{1,2}$ and $b_{1,2}$, is denoted by $\chi^2_{min}$.
In order to define our allowed parameter space, we keep only points which satisfy the following condition:
\be
\label{cl}
\Delta\chi^2=\chi^2-\chi^2_{min}\le 9.49(13.28)
\ee
where the value 9.49(13.28) corresponds to a 95(99)\% Confidence Level (CL) for
a $\chi^2$ with four degrees of freedom (dof). To better visualize the
allowed regions of the parameter space, we will project the four-dimensional
space into different planes. In this way, we will display the 95(99)\% CL EWPT
bounds on different selected pairs of free parameters.

\begin{figure}[!htbp]
\begin{center}
\unitlength1.0cm
\begin{picture}(7,4)
\put(-5.6,-4){\epsfig{file=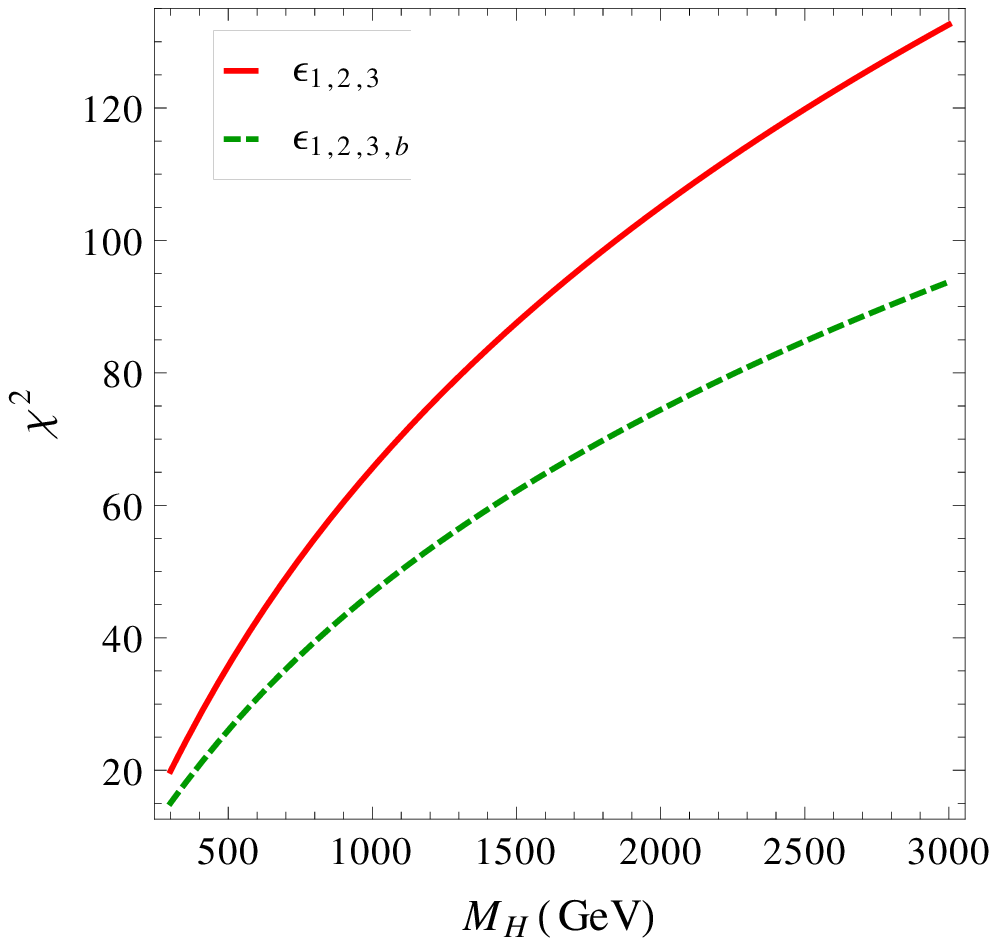,width=7.5cm}}
\put(3.5,-4){\epsfig{file=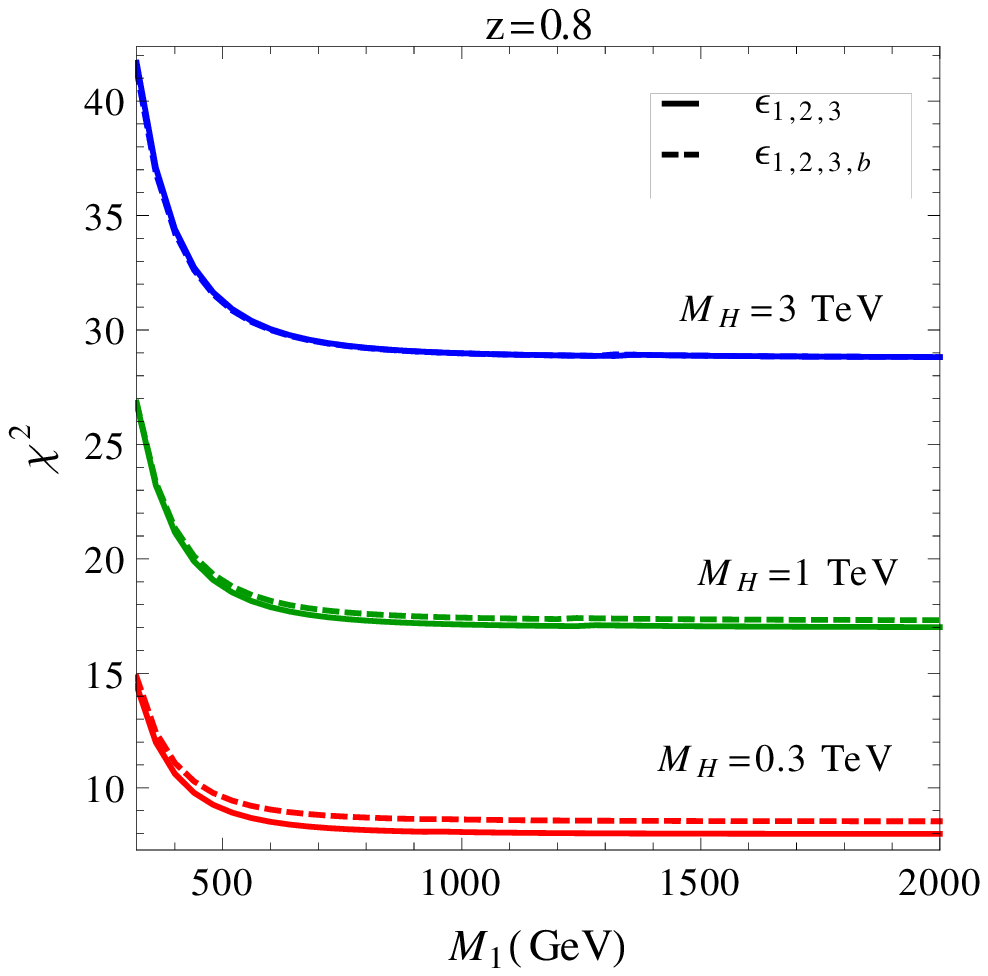,width=7.5cm}}
\end{picture}
\end{center}
\vskip 4.cm
\caption{Left: $\chi^2$-function versus the Higgs mass, $M_H$, within the SM.
The solid line comes from the correlated $\epsilon_{1,2,3}$ analysis, the
dashed one includes also the correlated $\varepsilon_b$ parameter.
Right: $\chi^2$-function versus the bare mass $M_1$, within the 4-site
Higgsless model at fixed $z=M_1/M_2=0.8$, after minimizing over the two
remaining free parameters, $b_{1,2}$. The solid line comes from the correlated
$\epsilon_{1,2,3}$ analysis, the dashed one includes also the correlated
$\varepsilon_b$ parameter. From bottom to top, the three sets of curves
correspond to the following three values of the $M_H$ parameter:
$M_H$=0.3, 1, 3 TeV.}
\label{chi2min1}
\end{figure}

But, before doing that, let us first discuss the statistical concept of
goodness-of-fit, which describes how well a theoretical model fits a set of
measurements. Qualitative arguments suggest that it can be summarized by the
condition $\chi^2_{min}\simeq$ dof. In Fig.~\ref{chi2min1}, we compare the
goodness-of-fit of the 4-site Higgsless model to electroweak precision
data expressed in terms of the $\varepsilon_i$ parameters (right plot) with
the analogous goodness-of-fit of the Standard Model (left plot).
In the right plot of Fig.~\ref{chi2min1}, we fix $z=M_1/M_2=0.8$,
and show how the $\chi^2$-function varies with $M_1$ once minimizing
over the two remaining $b_{1,2}$ free parameters. The solid lines
correspond to the correlated $\varepsilon_{1,2,3}$ analysis. The
dashed curves include also $\varepsilon_b$. Fig.~\ref{chi2min1}
clearly shows that $\varepsilon_b$ does not give a relevant
contribution to the 4-site Higgsless model test, and justifies our
choice to neglect it from now on, also, the $\varepsilon_b$
measurement is poorly correlated to the others (see
Eq.~(\ref{eq:LEP1})). From top to bottom, the three solid lines give
the $\chi^2$ function for three different values of the $M_H$
parameter in Eq.~(\ref{eps123l}): $M_H=3, 1, 0.3$ TeV respectively.
Independently on the value of $M_H$, the $\chi^2$ function is almost
flat in the $M_{1,2}$ mass parameters, except at very low bare
masses where it rapidly increases. All $z$ values share the same
feature. Thus, there is not a clear minimum $\chi^2$ in the
$M_{1,2}$ masses. The second information displayed in
Fig.~\ref{chi2min1} is the strong dependence of the $\chi^2$
function on the $M_H$ parameter. The $\chi^2$ increase with $M_H$
reflects the well known conflict between cut-off scale and new
physics content. The $\chi^2$ values obtained within the 4-site
Higgsless model can be compared with the SM $\chi^2$ for the same
$M_H$ values. The SM $\chi^2$ function versus $M_H$ is shown in the
left plot of Fig.~\ref{chi2min1}. In this way, the balance between
$M_H$ dependent terms and new physics contributions to the
$\varepsilon_i$ parameters is evident. The dramatic growth of the SM
$\chi^2$ function with increasing the $M_H$ parameter is largely
compensated by the new physics content predicted by the 4-site
Higgsless model. This shows that, despite the fact that Higgsless
models are characterized by large minimum $\chi^2$ values thus
failing the goodness-of-fit thumb rule $\chi^2_{min}/\mbox{dof}\le
1$, they succeed in curing the non-linear $\sigma$-model and
represent a
 viable alternative to the SM with a few hundred Higgs mass ($\chi^2_{SM}(M_H=0.3\,\mbox{TeV})\sim\chi^2_{min}(M_H=1\,\mbox{TeV})$).

\section{Mass spectrum and couplings of the extra $W_{1,2}^\pm$ and $Z_{1,2}$ gauge bosons}
\label{miai_s}
In this section, we derive the EWPT bounds on mass spectrum and couplings of the extra $W_{1,2}^\pm$ and $Z_{1,2}$ gauge bosons to
 ordinary matter. The aim is giving a complete definition of the physical properties of the new vector resonances predicted by the
4-site Higgsless model, needed for any phenomenological analysis.

\subsection{Mass spectrum}
\begin{figure}[!htbp]
\begin{center}
\unitlength1.0cm
\begin{picture}(7,4)
\put(-5.6,-4){\epsfig{file=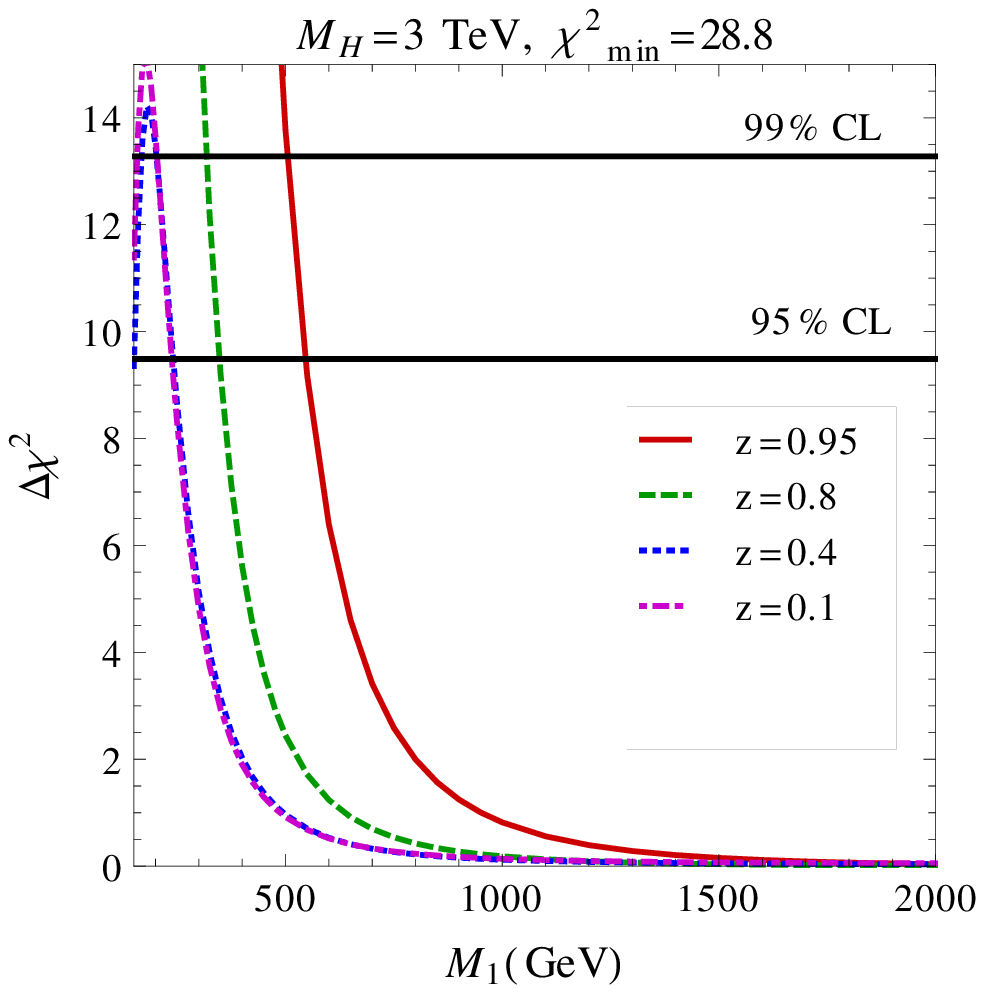,width=7.5cm}}
\put(3.5,-4){\epsfig{file=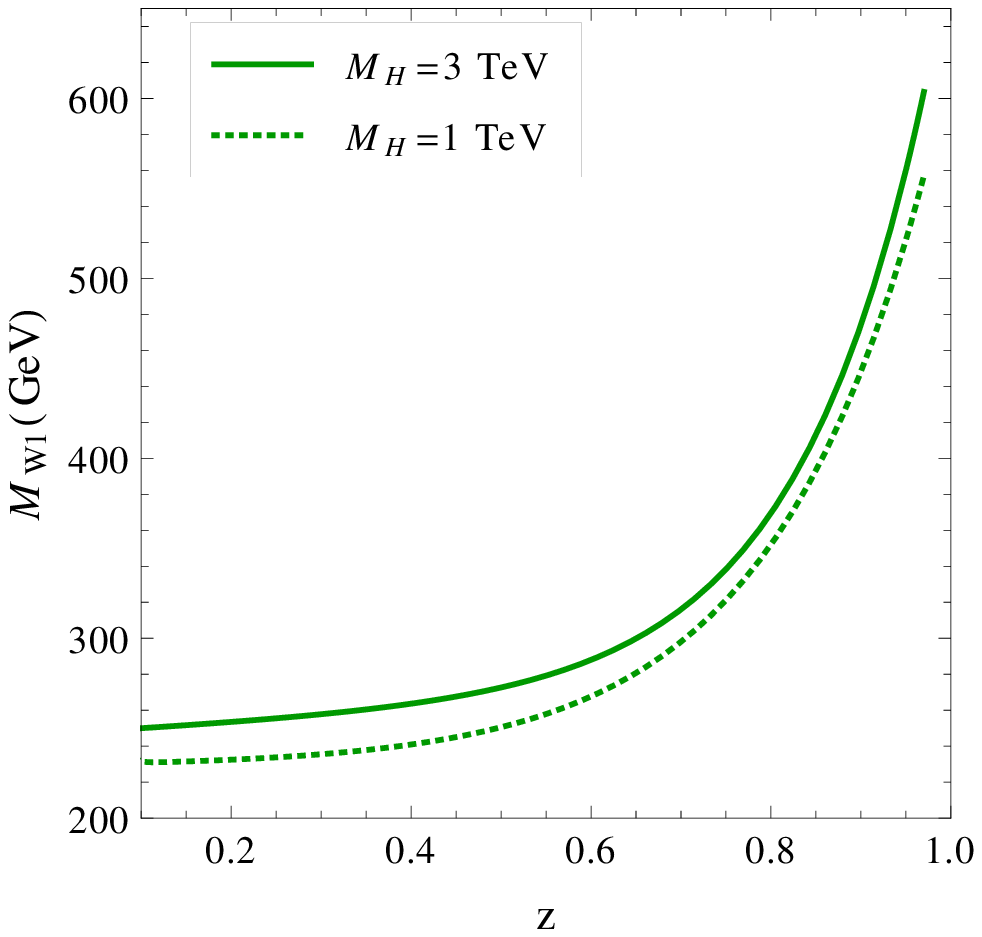,width=7.5cm}}
\end{picture}
\end{center}
\vskip 3.cm
\caption{Left: $\Delta\chi^2=\chi^2(z,M_1)-\chi^2_{min}$ function versus the bare mass parameter $M_1$, for four representative values of the
free parameter $z$ (see legend), once fixing the two remaining $b_{1,2}$ parameters at their optimal values. We choose $M_H$=3 TeV and use the
correlated $\epsilon_{1,2,3}$ analysis of Eq. (\ref{eq:LEP1}). The intersecting horizontal lines represent the 95\% and 99\% CL bound.
Right: Minimum mass of the lighter charged gauge boson $W_1^\pm$ as a
function of $z$. We fix $b_{1,2}$ to their optimal values, and consider two
values of the $M_H$ parameter: $M_H=$1 TeV (dashed line) and $M_H=$3 TeV
(solid line). We take the 95\% CL EWPT bound from the left plot.}
\label{deltachi2}
\end{figure}

A first information to be derived concerns the possible existence of a minimum allowed mass for the six extra resonances, $W^\pm_{1,2}$
and $Z_{1,2}$, predicted by the 4-site Higgsless model. In order to derive that, in the left plot of Fig.~\ref{deltachi2} we show
$\Delta\chi^2=\chi^2(z, M_1)-\chi^2_{min}$ as a function of $M_1$ for four representative $z$ values: $z$=0.1, 0.4, 0.8 and 0.95. We fix
$M_H$=3 TeV.
The function $\chi^2(z, M_1)$ is computed by minimizing over the two remaining $b_{1,2}$ parameters, while $\chi^2_{min}$ denotes the minimum
$\chi^2$ value over all four free parameters of the model. We use the correlated $\epsilon_{1,2,3}$ analysis of Eq. (\ref{eq:LEP1}).
The intersection of the above mentioned four curves with the solid horizontal lines gives the 95$\%$ and 99$\%$ CL lower bound on
the bare mass of the lighter extra gauge boson, $M_1$, according to Eq.~\refeq{cl}.
We now need to translate such a value into the minimum allowed physical mass for $W^\pm_{1,2}$ and $Z_{1,2}$ gauge bosons, taking into account
the corrections to the bare mass parameter discussed in the previous section. In the right plot of
Fig.~\ref{deltachi2}, we display the 95$\%$ CL contour in the ($z,M_{W1}$) plane for the two reference values of the $M_H$ parameter: $M_H$=1
and 3 TeV.
As one can see, the increase in $M_H$ gives a minor effect, shifting the minimum allowed mass by roughly 50 GeV, independently on $z$.

\subsection{$W_{1,2}^\pm$ and $Z_{1,2}$ gauge boson couplings to SM fermions}

\begin{figure}[!htbp]
\begin{center}
\unitlength1.0cm
\begin{picture}(7,4)
\put(-5.6,-4){\epsfig{file=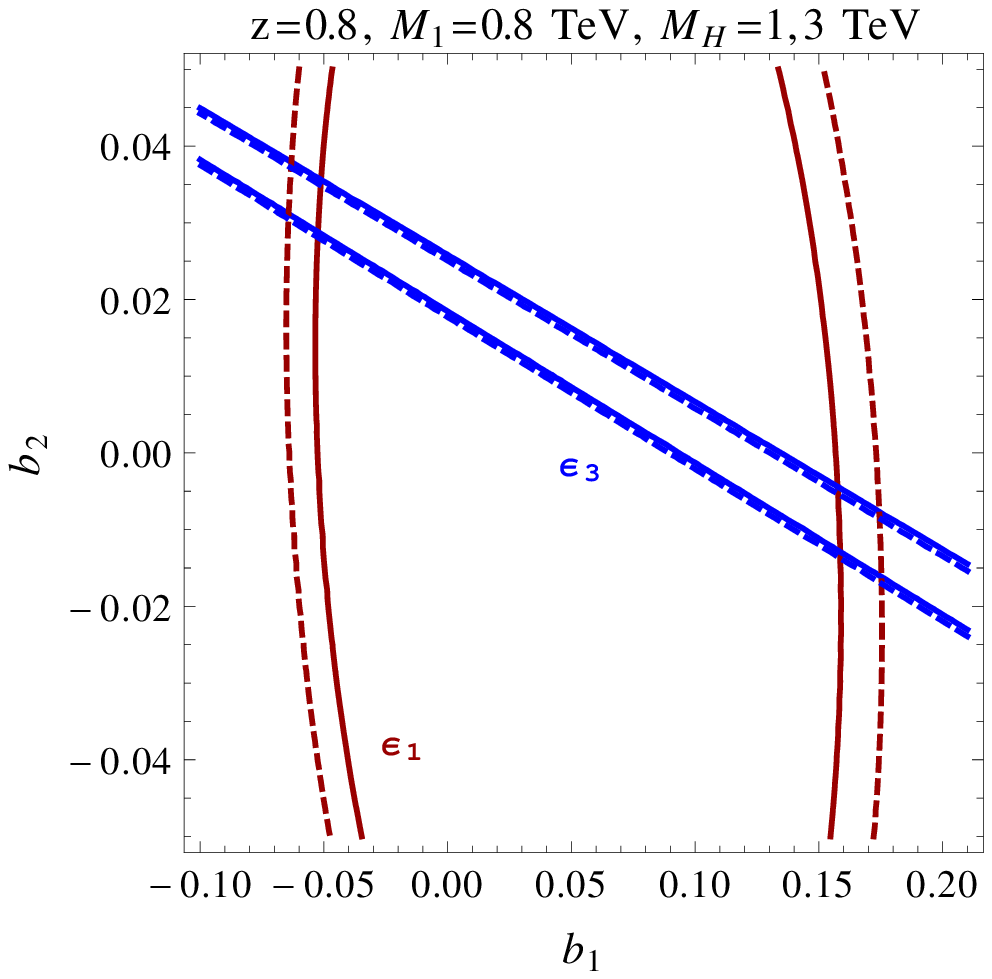,width=7.5cm}}
\put(3.5,-4){\epsfig{file=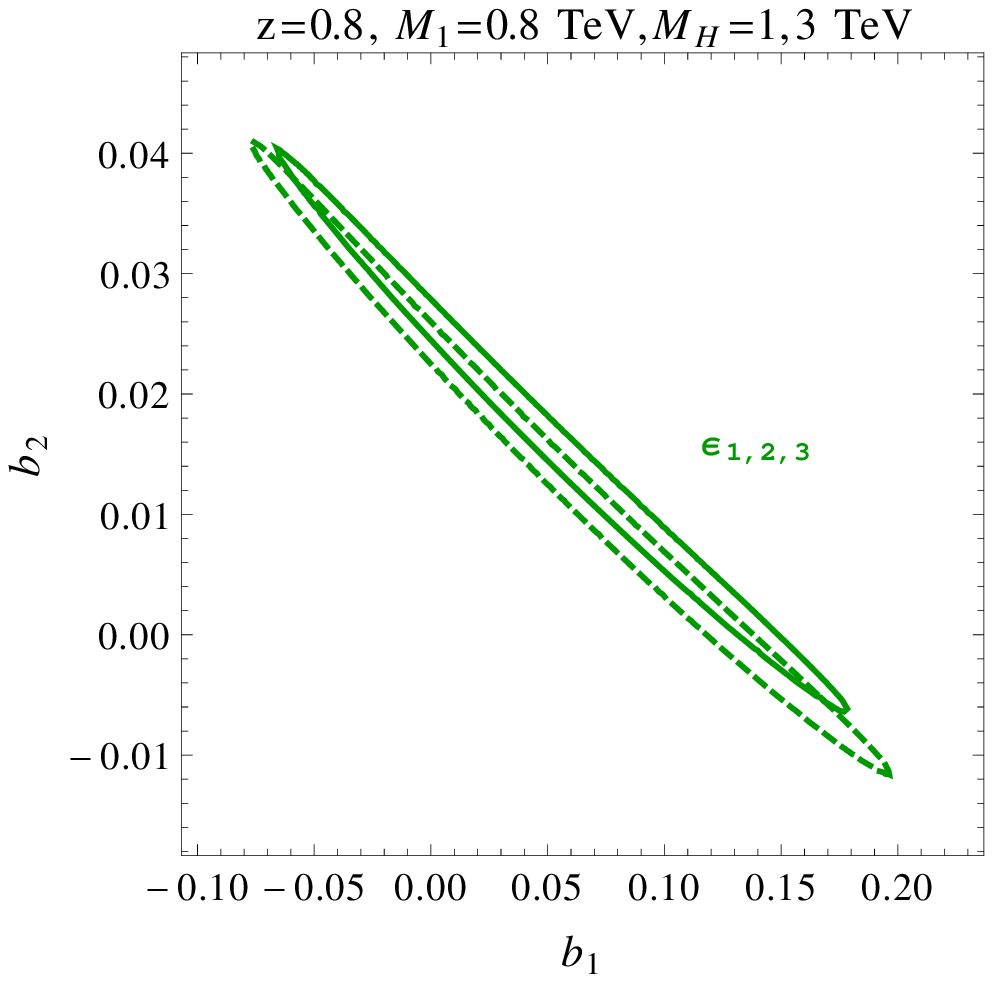,width=7.3cm}}
\end{picture}
\end{center}
\vskip 3.cm
\caption{Left: 95\% CL EWPT bounds in the $b_1,b_2$ plane at fixed $z$=0.8 and $M_1$=800 GeV (i.e. $M_2$= 1 TeV) from the individual
$\varepsilon_{1,3}$ parameters. The dashed lines correspond to $M_H$=1 TeV, the solid ones to
$M_H$=3 TeV. Right: 95\% CL EWPT bound on the $b_1,b_2$ plane at fixed $z$=0.8 from the fully correlated $\epsilon_{1,2,3}$ analysis.}
\label{b1b208}
\end{figure}

\begin{figure}[!htbp]
\begin{center}
\unitlength1.0cm
\begin{picture}(7,4)
\put(-5.6,-4){\epsfig{file=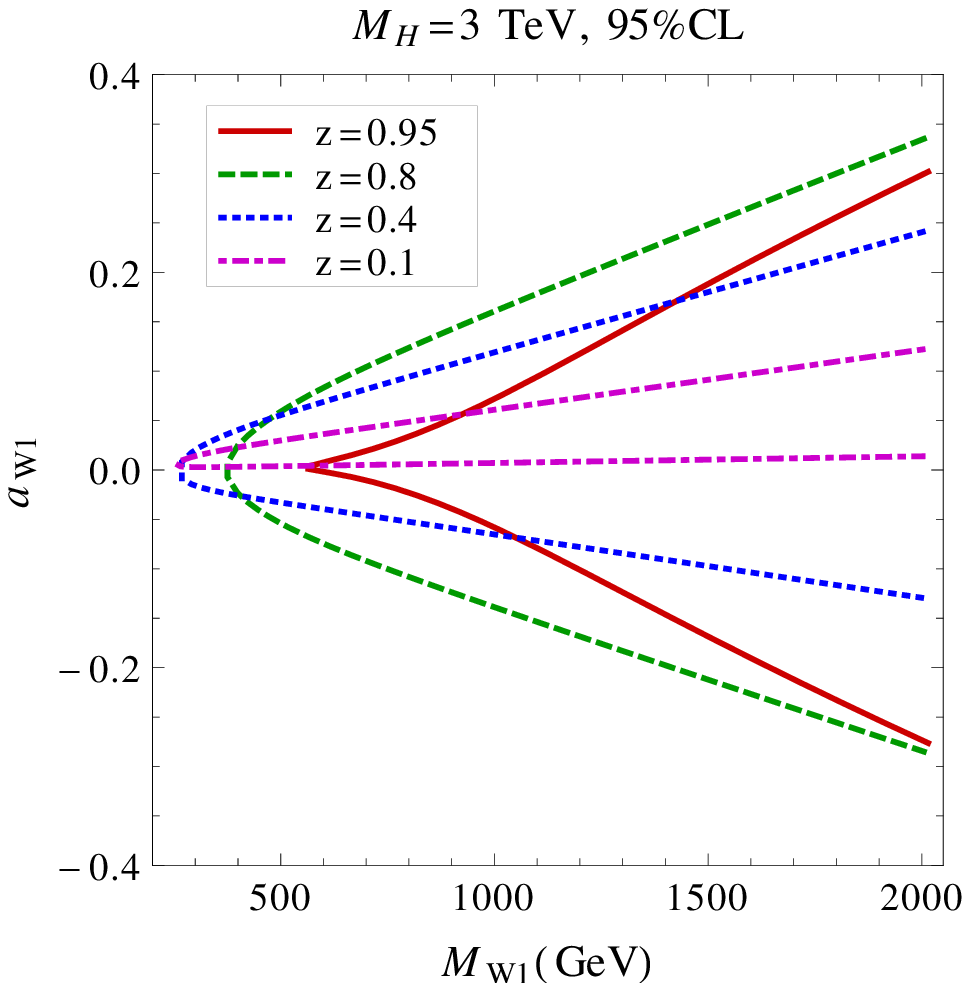,width=7.5cm}}
\put(3.5,-4){\epsfig{file=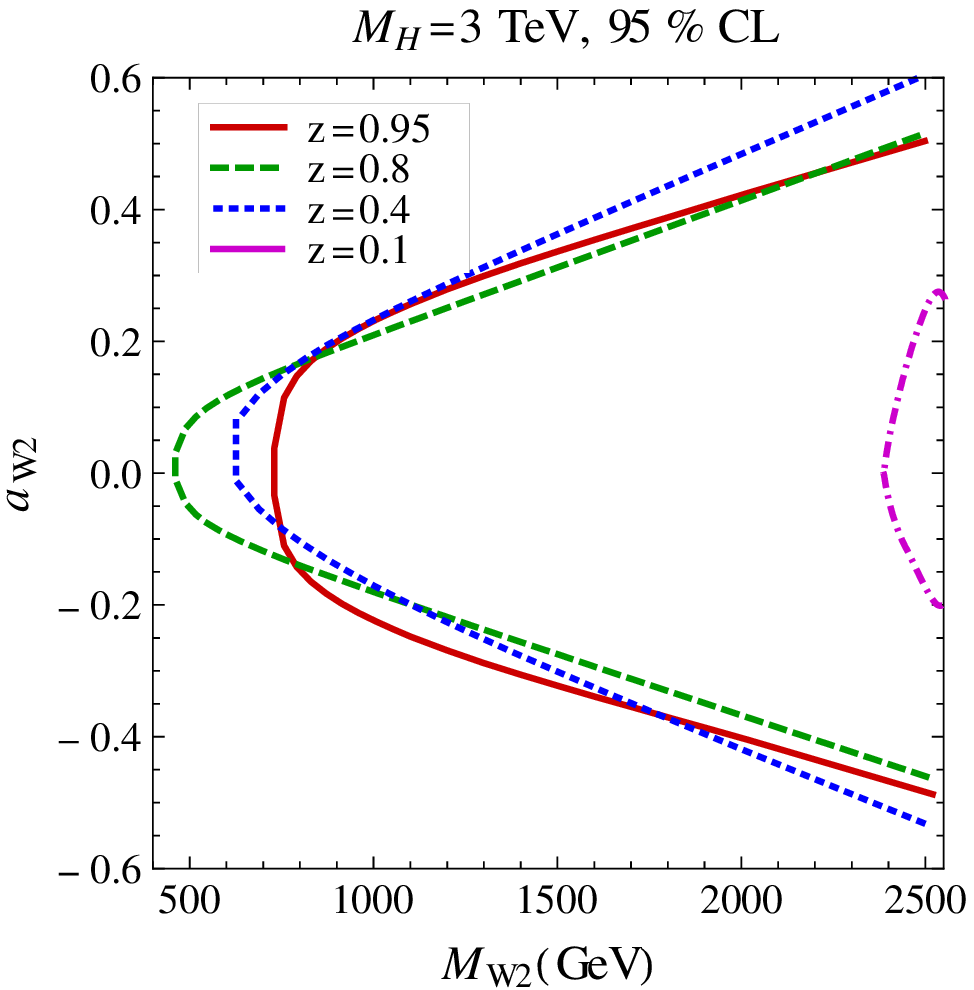,width=7.5cm}}
\end{picture}
\end{center}
\vskip 3.cm
\caption{Left: 95\% CL EWPT bound in the parameter space given in terms of physical mass, $M_{W1}$, and physical coupling between the
lighter extra gauge boson and SM fermions, $a_{W1}$ (see Eq.~\ref{aw}). We fix $M_H$=3 TeV, and consider four different $z$ values: $z$=0.1,
 0.4, 0.8 and 0.95
(see legend for linestyle code). Right: same for the heavier extra gauge boson $W_2^\pm$.}
\label{aw2b}
\end{figure}

In this section, we extract the EWPT bounds on the physical
couplings of the extra gauge bosons with ordinary matter. To this
aim, we start deriving the EWPT constraints on the two free
parameters of the 4-site Higgsless model, $b_{1,2}$, which represent
the bare direct boson-fermion gauge couplings. We project the
$\chi^2$ condition given in Eq.~\refeq{cl} on the $b_1,b_2$ plane at
fixed values of the two remaining parameters: $z$=0.8 and $M_1$=0.8
TeV (i.e. $M_2$=1 TeV). The results are shown in Fig.~\ref{b1b208}.
In the left plot, we display the 95$\%$ CL contour plot from the two
$\varepsilon_1$ and $\varepsilon_3$ parameters separately,
extracting their individual contributions from Eq.~\refeq{cl}. There
are two main information contained here. First, one can see that
while $\varepsilon_3$ forces the two $b_1,b_2$ couplings to be
almost linearly dependent, $\varepsilon_1$ imposes strong
constraints on their magnitude. The two $\varepsilon_{1,3}$
parameters play both an important role. Hence, oppositely to what
commonly done in the literature where the $\varepsilon_1$ tree level
contribution is neglected, it is mandatory to consider them jointly
when deriving the physical properties of the extra gauge bosons
predicted by Higgsless models. The second information concerns the
effect of the $M_H$ parameter. As clearly shown, it slightly affects
only the $b_1,b_2$ contour coming from the $\varepsilon_1$
parameter.
In the right plot of Fig.~\ref{b1b208}, we show the 95\% CL bound on the $b_1,b_2$ plane coming from the fully correlated
$\varepsilon_{1,2,3}$ analysis of Eq.~\refeq{cl}. As expected, the correlation shrinks the allowed $b_1,b_2$ area compared to the naive
uncorrelated $\varepsilon_{1,3}$ overlapping strip. Nevertheless, relatively sizable values for the bare direct couplings, $b_{1,2}$ are
allowed by EWPT. These values are mildly affected by the choice of $M_H$. In the following, we fix $M_H$=3 TeV.
The above mentioned results can be translated into direct limits on the physical couplings of the new vector bosons to SM fermions. In
 Fig.~\ref{aw2b}, we focus on the charged gauge sector, and we plot the 95$\%$ CL EWPT bounds in the
physical mass-coupling plane for the lighter (left plot) and heavier (right plot) extra vector bosons, $W_{1,2}^\pm$. We choose four
representative values for the $z$ free parameter: $z$=0.1, 0.4, 0.8 and 0.95.
The mass range is limited by the minimum mass previously discussed, and the
upper bound coming from the perturbative unitarity requirement (see Ref. \cite{Accomando:2008jh,Accomando:2008dm,Accomando:2010ir}
for details).
We should also notice that the signs of the physical fermion-boson couplings
are completely arbitrary and physically irrelevant (of course the couplings
of the different types of fermions to the same neutral boson are not
independent though). However, in the $b_1, b_2$ regions allowed by EWPT,
there is an almost two-fold degeneracy in the value
of each of the couplings. Two such points in the $b_1, b_2$ plane, for a given physical coupling, are not exactly equivalent as the other
fermion couplings would not generally be the same. Therefore we chose to give different signs to the physical couplings depending on which
 side of the allowed parameter-space they correspond to.
Fig.~\ref{aw2b} shows that the $z$ dependence is quite strong. For high-intermediate $z$ values, the allowed portion of the parameter space
 is large, and accommodates large values of the gauge couplings to SM fermions.
With decreasing $z$, the gauge boson-fermion
couplings get drastically reduced, approaching the almost fermiophobic scenario in the limit where $z$ tends to zero (in this case of course
 the heavier gauge boson decouples, and one recover the minimal 3-site Higgsless model with only $W_1^\pm$ and $Z_1$).
\begin{figure}[!htbp]
\begin{center}
\unitlength1.0cm
\begin{picture}(7,4)
\put(-5.6,-4){\epsfig{file=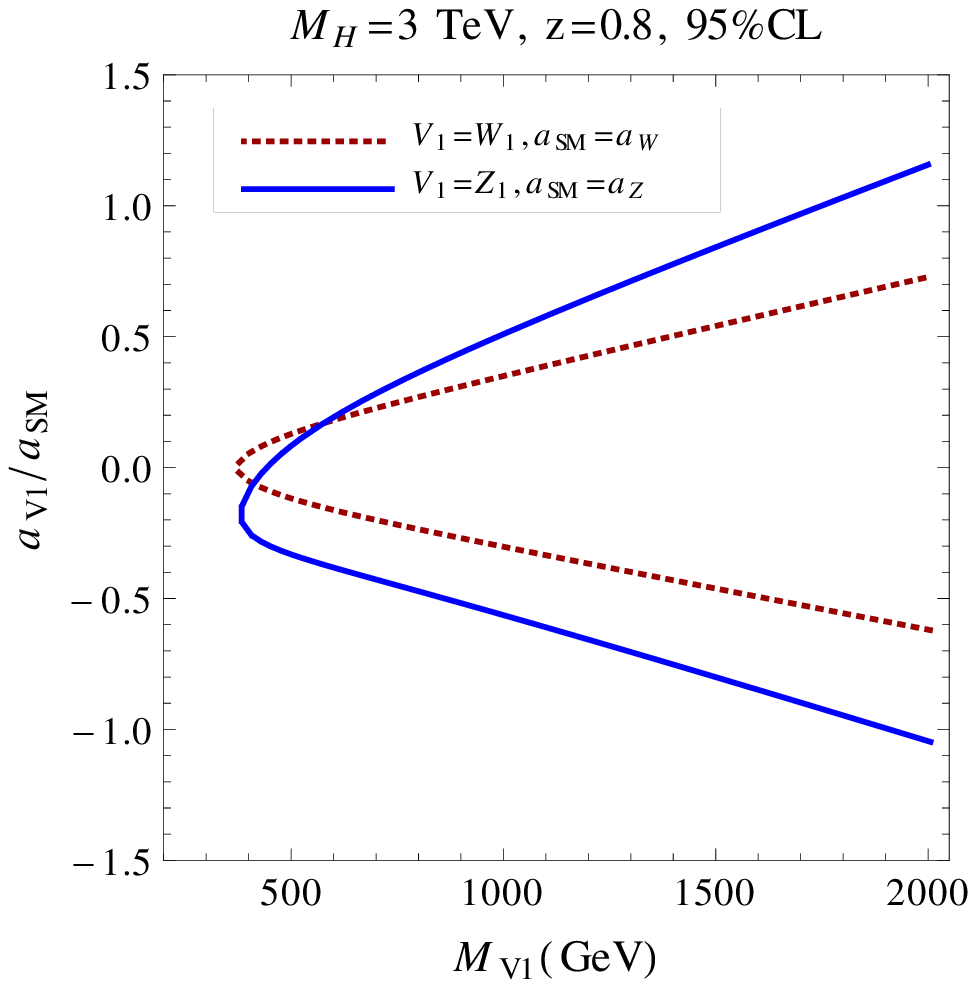,width=8cm}}
\put(3.5,-4){\epsfig{file=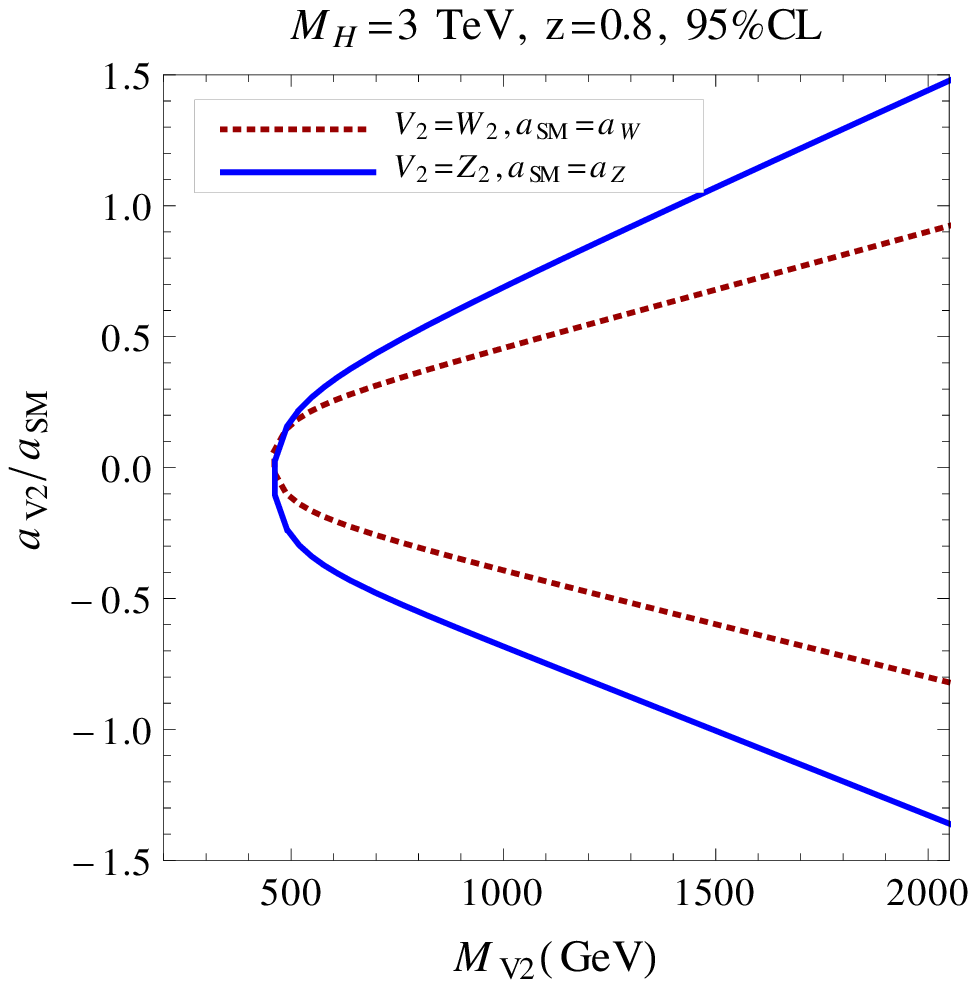,width=8cm}}
\end{picture}
\end{center}
\vskip 3.cm
\caption{Left: The solid line represents the 95\% CL EWPT bound in the physical mass-coupling plane for the lighter neutral gauge boson
$Z_1$, at fixed $z$=0.8. The mass is denoted by $M_{Z1}$. For the coupling, we choose as reference the gauge boson coupling to SM left-handed
 electrons, normalized to the corresponding SM one, $a_{Z1}/a_Z$. As comparison, the dashed line gives the parameter space of the lighter
charged gauge boson. Right: same for the heavier resonances $Z_2$ and $W_2^\pm$.}
\label{fig_ratios}
\end{figure}
For sake of completeness, in Fig.~\ref{fig_ratios} we show also the 95\% CL EWPT bounds in the mass-coupling plane for the neutral gauge
sector. We choose as reference the couplings between extra neutral gauge bosons and SM left-handed electrons, $a_{Z_i}=a_{Zi}^L(e)$ with
i=1,2. We fix $z$=0.8. The left plot (solid line) gives the parameter space for the lighter neutral gauge boson, $Z_1$, where this time
the gauge coupling is normalized to the corresponding SM one ($a_Z=a_Z^L(e)$). As comparison, the lighter charged gauge boson parameter
space is also shown (dashed line). Neutral and charged gauge couplings to
ordinary matter are comparable in size. Moreover, they can be of the same
order of magnitude than the corresponding SM ones. Analogously, the right
plot of Fig.~\ref{fig_ratios} shows that the same is true for the heavier
extra gauge boson, $Z_2$. In this case, the neutral gauge couplings can be even bigger than the SM ones up to a factor 1.5.
Finally, let us notice that the gauge couplings of the heavier resonances
are stronger than those of the lighter ones. This is a peculiar feature of the 4-site Higgsless model, and can have important phenomenological
 consequences.
If realized in nature, the heavier bosons could indeed produce more events than the lighter ones.

Summarizing, the new exact tree level computation of the EWPT bounds on the
4-site Higgsless model shows that the surviving parameter space is quite large.
Oppositely to the minimal 3-site Higgsless model, which is strongly constrained to be almost fermiophobic by EWPT as we will discuss in
 Sect.\ref{3-site_s}, its next-to-minimal 4-site extension can accommodate sizeable couplings between extra resonances and SM fermions.
The 4-site model, other than better describing the extra dimensional content of Higgsless theories characterized by the presence of multiple
resonances, has thus the potential of being detected during the early stage
of the LHC experiment in the Drell-Yan channel.

\section{Approximate versus Exact solution}
\label{App_vs_Exac}

In Refs. \cite{Accomando:2008jh,Accomando:2008dm}, we computed the
$\varepsilon_{1,2,3}$ parameters via a perturbative expansion in $x=e/g_1$, where $e$ is the electric charge and $g_1$ the extra gauge
 coupling. We calculated all terms up to the second order, $\mathcal O(x^2)$,
keeping the full
content in the two direct gauge boson-fermion couplings or delocalization parameters, $b_{1,2}$. For completeness, key steps of the
 procedure and approximate expressions for the $\varepsilon_{1,2,3}$ parameters are summarized in Appendix \ref{appA}.
\begin{figure}[!htbp]
\begin{center}
\unitlength1.0cm
\begin{picture}(7,4)
\put(-1,-4){\epsfig{file=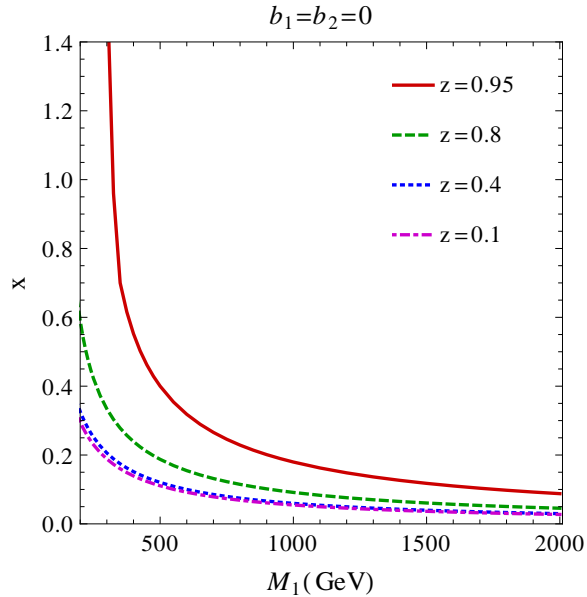,width=8cm}}
\end{picture}
\end{center}
\vskip 3.cm
\caption{Expansion parameter, $x=e/g_1$, as a function of the bare mass,
$M_1$, for $z$=0.1, 0.4, 0.8, and 0.95. We fix $b_{1,2}=0$.}
\label{fig_g1m1}
\end{figure}
In order to analyze the validity domain of the $\mathcal
O(e^2/g_1^2)$ approximation, in Fig.~\ref{fig_g1m1} we plot the
expansion parameter $x$ as a function of the bare mass $M_1$ for
four values of the free $z$ parameter: $z$=0.1, 0.4, 0.8, and 0.95.
While at large masses ($M_1\gtrsim1$ TeV) the neglected higher order
terms are expected not to exceed the permil level, they become more
and more important with decreasing $M_1$. Also, they are not
negligible approaching the limit $z\rightarrow 1$.
 This qualitatively indicates that, the series expansion breaks down for low masses ($M_1\lesssim1$ TeV) and high $z$ values ($z\sim1$).
In order to explore these regions, using the exact numerical calculation of the
$\varepsilon_i$ parameters is mandatory.
\begin{figure}[!htbp]
\begin{center}
\unitlength1.0cm
\begin{picture}(7,4)
\put(-5.6,-4){\epsfig{file=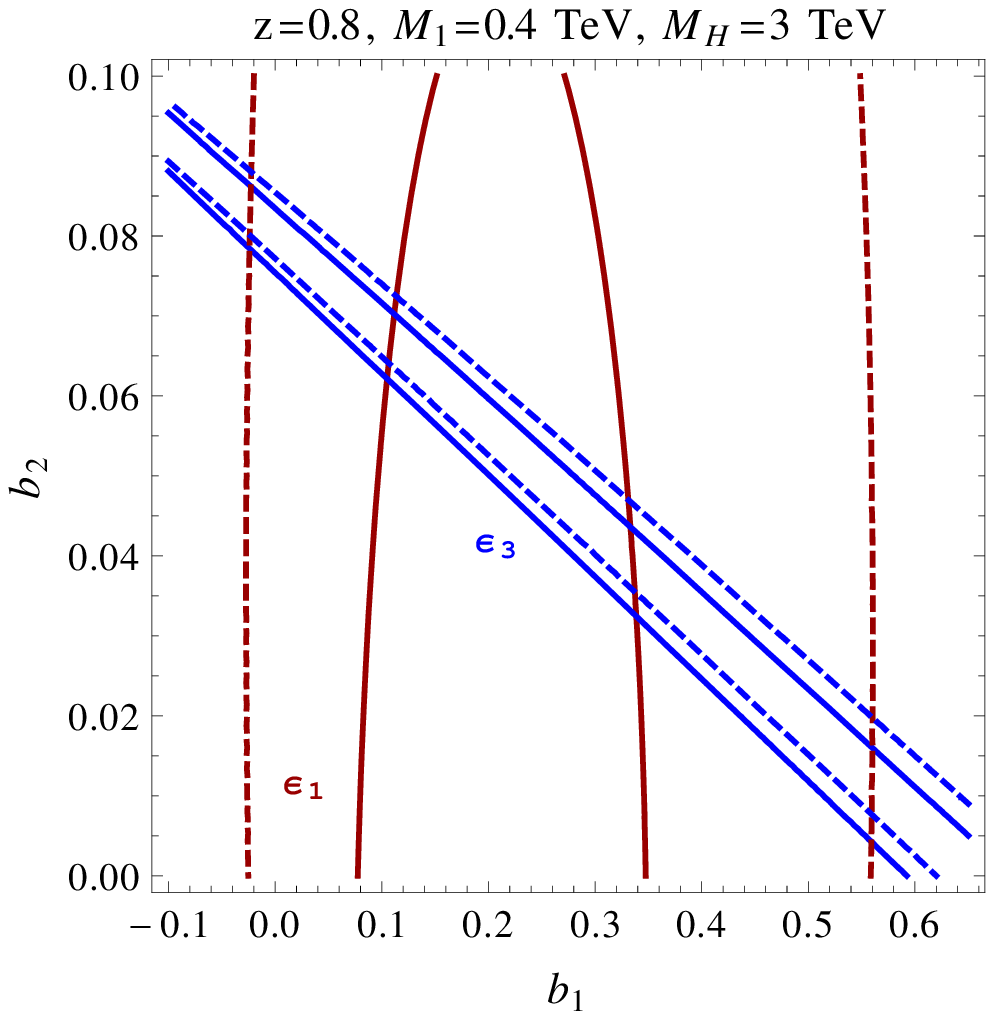,width=7.5cm}}
\put(3.5,-4){\epsfig{file=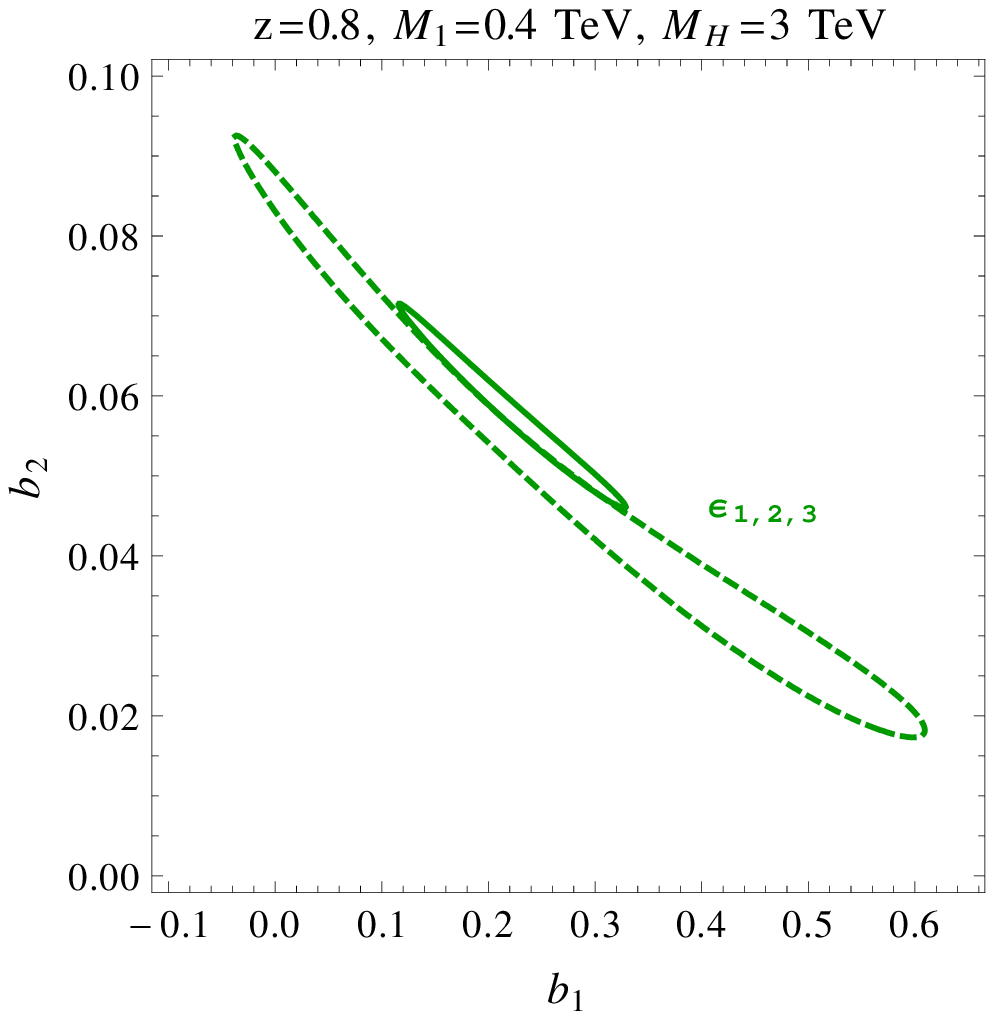,width=7.5cm}}
\end{picture}
\end{center}
\vskip 3.cm
\caption{Left: Comparison between exact (solid line) and $\mathcal O(x^2)$
approximate (dashed line) 95\% CL EWPT bound on the $b_1,b_2$ plane for
$z$=0.8 and $M_1$=0.4 TeV. We use the individual $\varepsilon_1$ and
$\varepsilon_3$ contributions to Eq.~(\refeq{cl}).
Right: Same, but employing the full correlated $\varepsilon_{1,2,3}$ analysis.}
\label{Fapproximated}
\end{figure}
More quantitatively, in Fig.~\ref{Fapproximated} we compare approximate and exact 95\% CL EWPT bounds in the $b_1,b_2$ plane for $z$=0.8
and $M_1$=400 GeV. In the left plot, we consider the individual $\varepsilon_1$ and $\varepsilon_3$ contributions to Eq.~(\ref{cl}) separately.
 And for each $\varepsilon_i$ (i=1,3) we show two progressive computational steps: the approximate results which take into account terms up
 to $\mathcal O(x^2)$ including the complete
$b_{1,2}$ content (dashed curves), and the exact numerical calculation at all orders (solid curves). The parameter $\varepsilon_2$ doesn't
 give any contribution in the range shown. We see that the
$\mathcal O(x^2)$ result is in good agreement with the exact result
for $\varepsilon_3$, while it fails in describing $\varepsilon_1$.
In the latter case, the exact contour drastically differs from the
$\mathcal O(x^2)$ approximate one. As a reference, in Appendix A,
the $\mathcal O(x^2)$ expressions for the $\varepsilon_{1,2,3}$
parameters are reported. These cumbersome formulas are exact in
$b_{1,2}$. They would assume a much simpler form by performing
either a first or even a second-order expansion in the $b_{1,2}$
parameters as well. This further approximations are largely used in
the literature. However our finding is that some cancelations may
occur and there is very little control over the validity of the
expansion (in particular one shouldn't neglect $x^2 b_{1,2}$ terms),
so we did not expand in $b_{1,2}$ at all. In the right plot of
Fig.~\ref{Fapproximated}, we display the exact (solid line) and
$\mathcal O(x^2)$ approximate (dashed line) 95\% CL EWPT bound on
the $b_1,b_2$ plane for $z$=0.8 and $M_1$=400 GeV, taking into
account the correlated $\varepsilon_{1,2,3}$ analysis. The
difference between the two calculations is certainly remarkable.

\begin{figure}[!htbp]
\begin{center}
\unitlength1.0cm
\begin{picture}(7,3)
\put(-1,-4.5){\epsfig{file=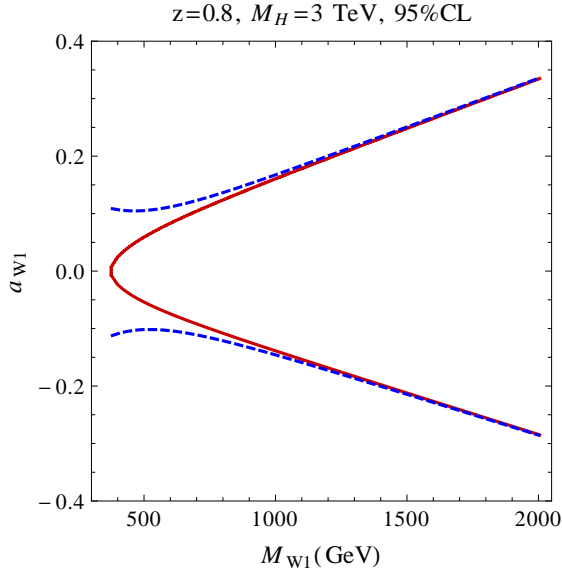,width=7.5cm}}
\end{picture}
\end{center}
\vskip 3.5cm
\caption{Comparison of the allowed region at 95\% CL in the plane ($M_{W_1},a_{W_1}$) between the approximated (blue dashed)
 and exact (red solid) solution for $z=0.8$.}
\label{Fapproximated2}
\end{figure}

To analyze the consequences of this behavior on the physical
quantities, in Fig.~\ref{Fapproximated2} we plot exact (solid line)
and $\mathcal O(x^2)$ truncated (dashed line) 95\% CL EWPT bounds in
the mass-coupling plane. We select the parameter space
($M_{W1},a_{W1}$) of the lighter charged gauge boson. The figure
confirms that for low masses, $M_1\lesssim 1$ TeV, the approximation
is not reliable anymore.

\section{3-Site Higgsless Model and EWPT bounds}
\label{3-site_s}
In this section, we specialize our results to the so called 3-site Higgsless
model. This model can be seen either as the minimal $K=1$ case of
deconstructed theories \cite{Casalbuoni:2005rs}, or as the BESS model with
$\alpha=1$
\cite{Casalbuoni:1985kq}. By imposing the LR symmetry in the gauge sector, it is a priori described by five parameters
($\tilde g,\tilde g',g_1,f_1,b_1$).
Fixing the gauge parameters $\tilde g,\tilde g',g_1$ in terms of the three SM inputs $e, G_F, M_Z$ as done before for the 4-site model,
the number of independent model parameters gets reduced to two: $M_1$ and $b_1$. These are the bare mass and the direct couplings to SM
fermions of the new gauge boson triplet, respectively. The 3-site model can be obtained from its 4-site extension by taking the limit
 $M_2\rightarrow\infty$ and $b_2=0$ with $M_1$ finite (or $z=b_2=0$ with $M_1$ finite).

\begin{figure}[!htbp]
\begin{center}
\unitlength1.0cm
\begin{picture}(7,4)
\put(-5.6,-4){\epsfig{file=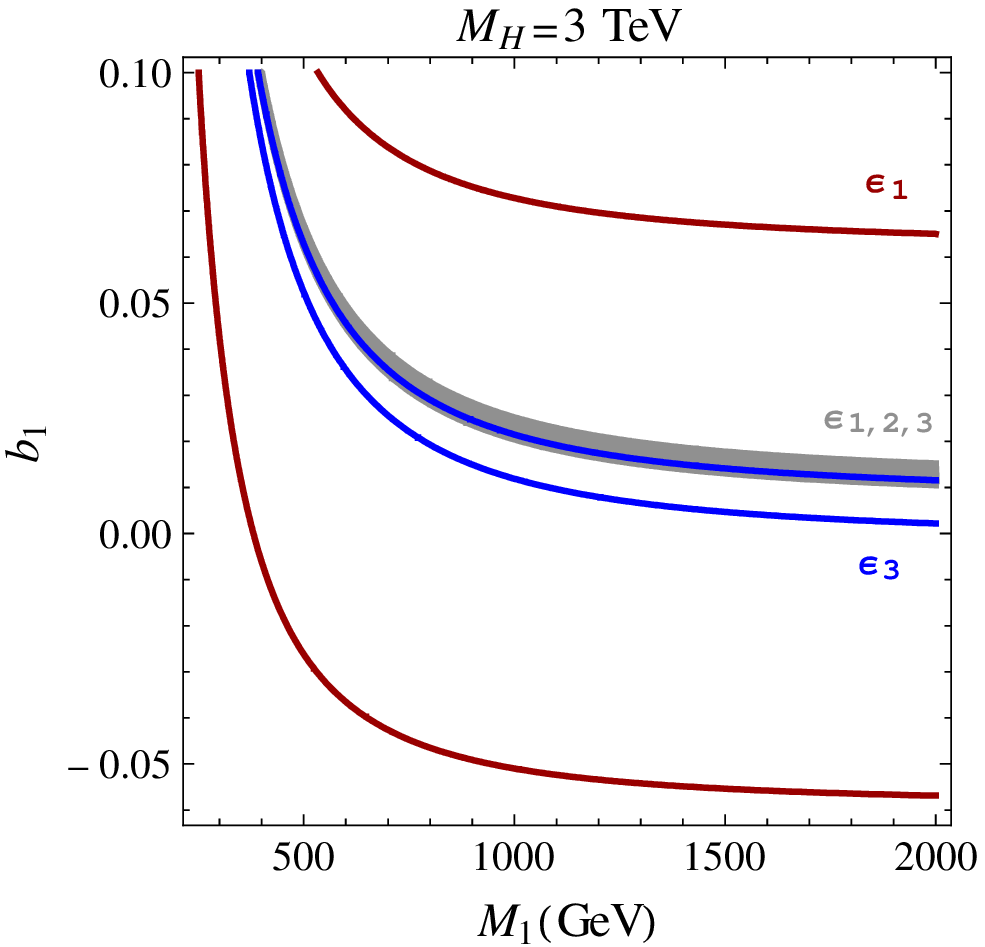,width=7.5cm}}
\put(3.5,-4){\epsfig{file=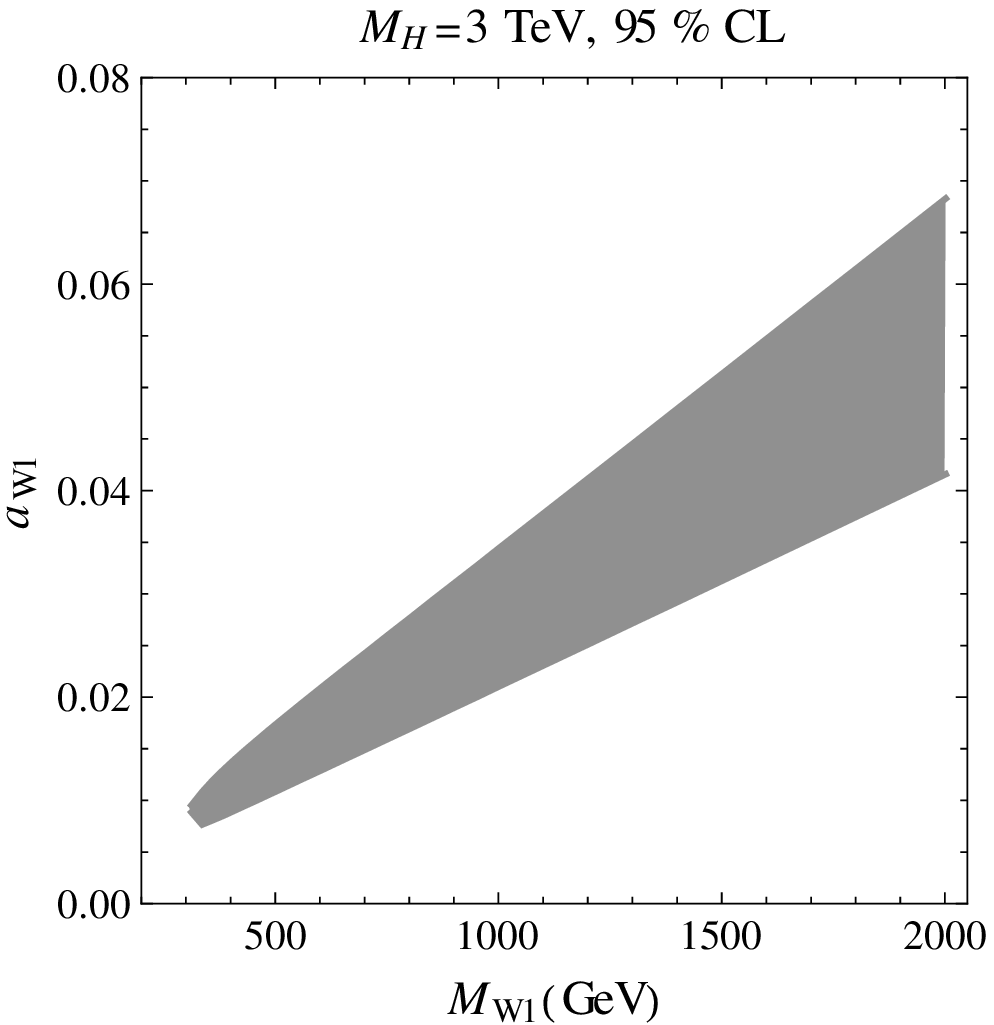,width=7.2cm}}
\end{picture}
\end{center}
\vskip 3.cm
\caption{Left: 95\% CL EWPT bound in the bare parameter plane $(M_1, b_1)$. The
wider region corresponds to the individual $\varepsilon_1$ contribution to
Eq.~\refeq{cl_3site}, the internal narrow region to the analogous $\varepsilon_3$ contribution. The fully correlated EWPT bound
is given by the gray region.
Right: 95\% EWPT bound in the physical mass-coupling plane ($M_{W1}, a_{W1}$) from the fully correlated $\varepsilon_{1,2,3}$ analysis.
 We fix $M_H=3$ TeV.}
\label{3-site}
\end{figure}

Analogously to what done for the 4-site model, we now derive the EWPT bounds on the 3-site Higgsless model, using an exact numerical
algorithm.
Since we have only two free parameters, $M_1, b_1$, the previous Eq.~\refeq{cl} must be replaced by
\be
\label{cl_3site}
\Delta\chi^2=\chi^2-\chi^2_{min}\le 5.99(9.21)
\ee
where the value 5.99(9.21) corresponds to a 95(99)\% CL for a $\chi^2$ with two degrees of freedom. As we can see from Fig.~\ref{deltachi2},
the
$\chi^2_{min}$ value is almost independent on $z$ and $M_1$ (for
$M_1\gtrsim 1$TeV). Thus, its value within the 3-site model is not expected to differ from that one we have in the 4-site model.
We indeed obtain
$\chi^2_{min}$=28.8 for $M_H$=3 TeV.
By applying Eq.~\refeq{cl_3site}, we derive the 95\% CL EWPT bound on the
$(M_1,b_1)$ plane, as shown in the left plot of Fig.~\ref{3-site}. The wider region represents the bound coming from the individual
$\varepsilon_1$ contribution to Eq.~\refeq{cl_3site}. The narrow internal area shows instead the analogous bound from $\varepsilon_3$.
In this case, the EWPT constraints on the model are completely dominated by the $\varepsilon_3$ parameter. The fully correlated EWPT
bound is the gray region and it is quite near to the one from $\varepsilon_3$.
These results are obtained via an exact numerical computation. Let us notice however that, within the 3-site Higgsless model,
the approximate double expansion in the $x=e/g_1$ and $b_1$ parameters works quite well. The following analytical expressions for the
$\varepsilon_{1,3}$ parameters
\be
\label{eps3-3site}
\varepsilon_1\simeq-\frac{b_1^2}{4},\quad
\varepsilon_3\simeq\frac{1}{2}(x^2-b_1)
\ee
are in excellent agreement with the exact result. Within the minimal 3-site model, one can thus safely apply a series expansion at second
 order in $x$ and first order in the delocalization parameter $b_1$, neglecting the
$\varepsilon_1$ contribution to the EWPT bounds.

The bounds on the two free parameters of the model can be translated into the
physical plane. In the right plot of Fig.~\ref{3-site}, we show the EWPT
constraint in the mass-coupling plane $(M_{W1},a_{W1})$. Here, $M_{W1}$
denotes the physical mass of the charged extra gauge boson, while $a_{W1}$
represents its coupling to SM fermions. We clearly see that the allowed
region is quite tiny, and the couplings are very small. Compared to
the 4-site model, while in the limit $z\rightarrow 0$ the 3-site model is recovered, couplings about five times larger
can be allowed for larger
values of $z$.
This feature has important phenomenological consequences. In their minimal representation (or 3-site),
deconstructed theories
appear to be observable only in production channels driven by triple and quartic gauge boson self couplings, the minimal scenario
almost fermiophobic. For that reason, the Higgsless literature is mostly focused on difficult multi-particle processes like vector boson
fusion and associated production of new gauge bosons with SM ones \cite{Birkedal:2004au,Belyaev:2007ss,He:2007ge}.
This is however the result of a crude theoretical approximation. Deconstructed theories can express their extra dimensional nature and
their physical properties in a more complete and realistic way via their next-to-minimal 4-site representation. This $K=2$ moose model,
even if truncated, gives in fact the first representation of the multi-resonance nature of extra dimensional theories characterized by KK
excitation towers. The addition of one more site, to the 3-site, changes completely the physical properties of the predicted extra gauge bosons.
The 4-site scenario is not fermiophobic anymore. It thus allows to search for evidence also in production processes driven by boson-fermion
 couplings.
In particular, within the 4-site model, the new resonances could be observed in the favoured Drell-Yan channel already with the data
collected in the early stage of the LHC experiment.
A first phenomenological analysis of the 4-site model at the LHC is given in \cite{Accomando:2008jh,Accomando:2008dm,Accomando:2010ir}

\section{Conclusions}
\label{conclusions} In this paper, we have derived the EWPT bounds
on the 4-site Higgsless model, which appears as the next-to-minimal
deconstructed $SU(2)$ theory in five dimensions
\cite{Csaki:2003dt,Agashe:2003zs,Csaki:2003zu,Barbieri:2003pr,Nomura:2003du,
Cacciapaglia:2004zv,Cacciapaglia:2004rb,Cacciapaglia:2004jz,Contino:2006nn,ArkaniHamed:1998rs,Randall:1999ee}.
The model is based on the $SU(2)_L\times SU(2)_1\times SU(2)_2\times
U(1)_Y$ gauge symmetry, and predicts four charged $W^\pm_{1,2}$ and
two neutral $Z_{1,2}$ extra gauge bosons. Its novelty, compared to
the minimal 3-site representation, consists in reconciling EWPT
bounds and unitarity constraints without imposing the extra vector
bosons to be fermiophobic (owing to the inclusion of direct
fermion-boson gauge couplings in addition to those ones coming from
usual mixing terms).

The phenomenology of the 4-site Higgsless model is controlled by
only four free parameters beyond the SM ones: the bare masses,
$M_{1,2}$, of lighter and heavier extra gauge boson triplets and
their bare direct couplings to SM fermions, $b_{1,2}$. In this
paper, we have performed a new analysis of the EWPT constraints on
the aforementioned 4-dimensional parameter space. We used the
$\varepsilon_{1,2,3}$ parametrization of the universal electroweak
radiative corrections to the precision observables measured by LEP,
SLD and TEVATRON experiments. We neglected the $\varepsilon_b$
effect, as it is weakly correlated to the other measurements and
also because it receives no contribution within the 4-site model
owing to universality in the fermionic sector.

The four main novelties of our analysis can be summarized as follows. We
computed for the first time the $\varepsilon_i$ (i=1,2,3) parameters at tree
level via a complete numerical algorithm, going beyond commonly used
analytical approximations. In addition, we have taken into account the full
correlation between their measurements, performing a well defined and
complete statistical analysis,
based on the minimum $\chi^2$ test. We furthermore studied the cutoff
dependence of the derived EWPT bounds, and discussed how well the 4-site
Higgsless model can reproduce experimental results. We have finally shown a
one-to-one comparison between the EWPT surviving parameter space, given in
terms of
physical mass and coupling of the first charged resonance ($M_{W1},a_{W1}$),
within the minimal (3-site) and next-to-minimal (4-site) deconstructed Higgsless
models.

Our findings are as follows. The popular approximations existing in
the literature cannot give a reliable description of masses and
couplings allowed by EWPT over the full parameter space. The
second-order expansion in the $x=e/g_1$ parameter, keeping the full
dependence on the direct gauge boson-fermion couplings $b_{1,2}$ as
reported in Appendix A, is indeed valid only beyond $\mathcal
O$(TeV) mass scales and for low-intermediate values of the ratio
$z=M_1/M_2$ between the bare masses of the two predicted gauge
triplets ($z\lesssim$ 0.8). The validity range is mainly constrained
by the $\varepsilon_1$ parameter, $\varepsilon_3$ being rather
stable under the series expansion in $x$. Further truncating this
$\mathcal O(x^2)$ expansion up to either first-order or even
second-order terms in the remaining $b_{1,2}$ parameters as well, as
commonly done in the literature, would worsen the goodness of the
approximation sensibly. Taking into account the complete
contribution from the delocalization parameters, $b_{1,2}$, is thus
mandatory in order to extract reliable EWPT bounds on the 4-site
model. This also implies that one should consider $\varepsilon_1$ on
the same footing as $\varepsilon_3$. Despite the fact that at
leading order in the three parameters $x, b_1, b_2$, they go like
$\varepsilon_1\simeq b_i^2$ and $\varepsilon_3\simeq b_i+x^2$, both
$\varepsilon_{1,3}$ play a strong role. The $\varepsilon_3$
parameter generates an almost linear relation between the gauge
couplings of lighter and heavier extra resonances with ordinary
matter, while $\varepsilon_1$ constrains their size.

The new complete calculation of the EWPT bounds presented in this
paper takes into account all $\varepsilon_i$ (i=1,2,3) parameters
with their full correlation. We have found that this has indeed a
significant effect in extracting the allowed parameter space, as
compared to previously used simple analysis. The cutoff dependence
of our results is instead rather mild. Its major effect appears in
the minimum $\chi^2$ value that one can obtain within the 4-site
model. This value rapidly increases with the cutoff.

All these combined effects determine the portion of the parameter
space which survives to EWPT. The four-dimensional parameter space
of the 4-site model can be expressed in terms of physical masses and
couplings to fermions of the extra gauge bosons. A first EWPT effect
is to put a lower bound on the mass spectrum. If we take the lighter
charged extra gauge boson $W_1^\pm$ as representative, we find
indeed that its minimum mass can range between $M_{W1}^{min}$=250
and 600 GeV for 0.1$<z<$0.95. The second important result is that,
even if bounded, the gauge couplings of the six extra gauge bosons
to ordinary matter can be of the same order of magnitude than the
corresponding SM ones. This is in contrast with the almost
fermiophobic scenario of the minimal 3-site representation of
Higgsless theories. The addition of one more site brings a drastic
change. The next-to-minimal 4-site extension can in fact express the
multi-resonance nature of extra-dimensional theories, characterized
by Kaluza-Klein excitation towers, and give a less constrained
description of the physical properties of the predicted extra gauge
bosons.

An immediate phenomenological consequence is that the Drell-Yan
production process becomes an open channel for the direct search of these
new resonances already during the present data collection by LHC and TEVATRON
experiments. A first phenomenological analysis
of the 4-site model was done in Ref. \cite{Accomando:2008dm} and refined in
Ref. \cite{Accomando:2010ir} with a focus on the neutral gauge sector
$Z_{1,2}$. A detailed study concerning exclusion at the TEVATRON and
discovery reach at the LHC is now under investigation.

\begin{acknowledgments}
E.A. and D.B.  acknowledge financial support from the NExT Institute
and SEPnet. The work of S.D.C., D.D. and L.F. is partly supported by
the Italian Ministero dell'Istruzione, dell'Universit\`a
e della Ricerca Scientifica, under the COFIN program (PRIN 2008).\\
\end{acknowledgments}

\appendix

\section{Approximated analytical expressions for
$\epsilon_1,\epsilon_2,\epsilon_3$}
\label{appA}
We collect here some analytical formulas, which are necessary to express the
predictions for the observables of the 4-site model in terms of physical quantities
and the new parameters.
All the following definitions are expressed in terms of the model
parameters $g_1,\gt,\tilde g',M_1,z=M_1/M_2,b_1,b_2$.
Let us start with the mass eigenvalues. At the order $(\tilde g/g_1)^2$ they are:
\be\label{Mzt}
M_W^2=\tilde{M_W}^2\left(1-\tilde
x^2z_{W}\right),
\quad M_Z^2=\tilde{M_Z}^2\left(1-\tilde
x^2z_{Z}\right) \ee
\be
M_{W_1}^2=M_1^2\left(1+\frac{\tilde x^2}{2}\right),\quad
M_{W_2}^2=\frac{M_1^2}{z^2}\left(1+\frac{\tilde x^2z^4}{2}\right) \ee
\be
M_{Z_1}^2=M_1^2\left(1+\frac{\tilde x^2}{2\tilde c^2}\right),\quad
M_{Z_2}^2=\frac{M_1^2}{z^2}\left(1+\frac{\tilde x^2 z^4}{2\tilde c^2}\right)
\ee
with
\be\label{1}
\tilde{M_W}^2=\frac{\tilde{x}^2}{2}(1-z^2)
M_1^2,\quad\tilde{M_Z}^2=\frac{\tilde{M_W}^2}{\tilde{c}^2},\quad
\tilde x=\frac{\tilde g}{g_1}\quad
\ee
\be\label{zz}
z_W=\frac{1}{2}(1+z^4),\quad z_Z=-2\tilde s^2+\frac{z_W}{\tilde c^2}
\ee
and $ \tan\tilde\theta\equiv \tilde s/\tilde c =\gtp/\gt$.
We recall also the  couplings of $A^\mu$, $Z^\mu$, $Z_{1,2}^\mu$, $W^\mu$, $W_{1,2}^\mu$,
 to fermions:
\bea
{\mathcal L}_{NC}&=&\bar{\psi} \gamma^\mu \left [- e \mathbf{Q}^f A_\mu
+ {a}_{Z}^f Z_\mu
+{a}_{Z_1}^f
Z_{1\mu}+ {a}_{Z_2}^fZ_{2\mu}  \right ]\psi\nn\\
{\mathcal L}_{CC}&=&\bar{\psi} \gamma^\mu T^-  \psi \left(a_W
W_{\mu}^+ +a_{W_1} W_{1\mu }^+ +a_{W_2} W_{2\mu }^+\right) + h.c.
\eea
where:
\bea\label{az}
e&=&\tilde g \tilde s\left(1-\tilde x^2 \tilde s^2\right)\nn\\
a_Z^f&=&-\frac{\tilde g}{\tilde
c}\left(1-\frac{b}{2}\right)\left(1+\tilde
x^2\left(-\frac{z_{Z}}{2}+z_{Zb}\right)\right)
\left[\mathbf{T_3
}^f-\tilde s^2\frac{1+\tilde x^2(\tilde c^2-\tilde
s^2-z_{Zb})}{\left(1-\frac{b}{2}\right)}\mathbf{Q}^f\right]
\nn\\
a_{Z_1}^f&=&-\frac{g_1}{\sqrt 2(1+b_+)}\left(b_+-\frac{\tilde x^2}{\tilde c^2}(1+z^n_1)\right)\mathbf{T_3}^f+
\frac{g_1\tilde x^2\tilde s^2}{\sqrt2\tilde c^2}\mathbf{Q}^f\nn\\
a_{Z_2}^f&=&-\frac{g_1}{\sqrt 2(1+b_+)}\left(b_--\frac{\tilde x^2z^2}{\tilde c^2}(1+z^n_2)\right)\mathbf{T_3}^f-
\frac{g_1\tilde x^2z^2\tilde s^2}{\sqrt2\tilde c^2}\mathbf{Q^f}
\eea
and
\bea\label{aw}
a_W&=&-\frac{\tilde g}{\sqrt 2}\left(1-\frac{b}{2}\right)\left(1+\tilde
x^2\left(-\frac{z_{W}}{2}+z_{Wb}\right)\right)\nn\\
a_{W_1}&=&-\frac{g_1}{2(1+b_+)}\left(b_+-\tilde x^2(1+z_1)\right)\nn\\
a_{W_2}&=&-\frac{g_1}{2(1+b_+)}\left(b_--\tilde x^2z^2(1+z_2)\right)
\eea
with  $z_Z$ and $z_W$ given in (\ref{zz}), and
\be
\label{zzb}
z_{Zb}=(1-z^2)\frac{b_+(\tilde
c^2-\tilde s^2)+b_- z^4}{2(2+b_++b_-z^2)\tilde c^2} ,\quad
z_{Wb}=(1-z^2)\frac{b_++b_- z^4}{2(2+b_++b_-z^2)} \ee
\be z_1=\frac{b_+}{4}+b_-\frac{z^2}{2(1-z^2)},\quad
z_2=z^2\left(\frac{b_-}{4}-b_+\frac{1}{2(1-z^2)}\right)
\ee
\be
z^n_1=\frac{b_+(1-4\tilde s^2)}{4}+b_-\frac{z^2 (\tilde c^2-\tilde s^2)}{2(1-z^2)},\quad
z^n_2=\frac{b_-z^2}{4}-b_+\frac{z^2-2\tilde s^2}{2(1-z^2)}
\ee
\be
b=\frac{b_+-b_-z^2}{1+b_+}\quad\quad b_\pm=b_1\pm b_2
\ee
Now we want to write all in terms of $e,M_Z,G_F$ and the 4-site parameters
$ z,b_1, b_2,M_1$ (at the order $\tilde x^2=(\tilde g/g_1)^2$). Let us start with $\tilde g$
\be\label{gte}
\tilde g =\frac e {\tilde s}\left(1+x^2\right)\Rightarrow\tilde
x=\frac x {\tilde s}\left(1+x^2\right),~~~~ x=\frac e g_1
\ee
By computing the Fermi constant  $G_F$ as:
 \be \frac{G_F}{\sqrt
2}=\frac{a_W^2}{4M_W^2}+\frac{a_{W_1}^2}{4M_{W_1}^2}+\frac{a_{W_2}^2}{4M_{W_2}^2}
\ee
we get:
\bea\label{GFl} \frac{G_F}{\sqrt 2}&=&
\frac{e^2}{8M_Z^2\tilde c^2 \tilde
s^2}\left[\left(1-\frac{b}{2}\right)^2+(1-z^2)\frac{b_+^2+z^2b_-^2}{4(1+b_+)^2}\right]\nn\\
&&+\frac{e^2 x^2}{8M_Z^2\tilde c^2 \tilde s^2}
\left[\left(1-\frac{b}{2}\right)^2\left(2-\frac{z_Z}{\tilde
s^2}+2\frac{z_{Wb}}{\tilde s^2}\right)+
\left(2-\frac{z_Z}{\tilde s^2}\right)(1-z^2)\frac{b_+^2+z^2b_-^2}{4(1+b_+)^2}\right]\nn\\
&&-\frac{e^2 x^2}{8M_Z^2\tilde c^2 \tilde s^2} \frac{1-z^2}{\tilde
s^2}\left[\frac{b_+^2+z^6 b_-^2}{8(1+b_+)^2}+\frac{b_+(1+z_1)+z^4
b_-(1+z_2)}{2(1+b_+)^2}\right]
\eea
Let us now define the Weinberg angle $\theta$ by \cite{Altarelli:1991zd}:
\be \frac{G_F}{\sqrt
2}=\frac{e^2}{8\s^2\c^2M_Z^2}
\label{theta}
\ee
with $\s=\sin\theta$.
So Eq.~(\ref{GFl}) and Eq.~(\ref{theta}) imply:
\bea
\frac{e^2}{8\s^2\c^2M_Z^2}&=&\frac{e^2}{8\tilde s^2\tilde
c^2M_Z^2}X+\frac{e^2}{8M_Z^2}x^2A\quad\Rightarrow\nn\\
\tilde s^2\tilde c^2&=&\c^2\s^2X(1+\s^2\c^2x^2A)
\eea
with
\be
X=\left(1-\frac{b}{2}\right)^2+\frac{\beta}{4}\quad\quad
\beta=(1-z^2)\frac{b_+^2+z^2b_-^2}{(1+b_+)^2}
\ee \bea
A&=&\frac{1}{\tilde s^2 \tilde c^2}\left[X\left(2-\frac{z_Z}{\tilde
s^2}\right)+ \left(1-\frac{b}{2}\right)^2\frac{2z_{Wb}}{\tilde
s^2}-\frac{B}{\tilde s^2}\right] \nn\\
B&=&\frac{1-z^2}{8(1+b_+)^2}\left[b_+^2+z^6b_-^2+4b_+(1+z_1)+4b_-z^4(1+z_2)\right]
\eea
Now we can solve this equation perturbatively in $x$;
at the $x^2$ order we get:
\bea\label{0order}
\tilde s^2&=&s_*^2+x^2\frac{\s^4\c^4}{\sqrt{1-s_{2\theta}^2X}}AX\nn\\
\eea
with:
\bea\label{star}
s_*^2&=&\frac12\left(1-\sqrt{1-s_{2\theta}^2X}\right)\nn\\
s_*^2c_*^2&=&\s^2\c^2X
\eea
with  $\tilde s$ and $\tilde c$ replaced by
 $s_*$ and $c_*$  in $A$ since it is multiplied by $x^2$. Namely, by using the zero order of Eq. (\ref{star}), we can rewrite the zero order expression for $A$ as:
\bea A\vert_{x=0}&=&\frac{1}{ s_*^2
c_*^2}\left[X\left(2-\frac{z_Z}{ s_*^2}\right)+
\left(1-\frac{b}{2}\right)^2\frac{2z_{Wb}}{s_*^2}-\frac{B}{s_*^2}\right]\nn\\
&=&\frac{1}{ \s^2  \c^2}\left[\left(2-\frac{z_Z}{ s_*^2}\right)+
\left(1-\frac{b}{2}\right)^2\frac{2z_{Wb}}{Xs_*^2}-\frac{B}{Xs_*^2}\right]
\eea
Then, from Eqs.~(\ref{gte}) and (\ref{Mzt}), we get :
\bea
g_1^2&=&
\frac{e^2(1-z^2)M_1^2}{2\tilde s^2\tilde c^2 M_Z^2}\left(1+ x^2\left(2-\frac{z_Z}{\tilde
s^2}\right)\right)\nn\\
&=&\frac{e^2(1-z^2)}{2 s_*^2c_*^2 M_Z^2}\left(1-\s^2\c^2x^2A\right)\left(1+
x^2\left(2-\frac{z_Z}{\tilde s^2}\right)\right)\nn\\
&=&\frac{e^2(1-z^2)}{2 \s^2\c^2
XM_Z^2}\left[1+x^2\left(-\left(1-\frac{b}{2}\right)^2\frac{2z_{Wb}}{Xs_*^2}+\frac{B}{Xs_*^2}\right)\right]\nn\\
\tilde
g&=&\frac{e}{s_*}\left[1+x^2\left(1-\frac{\s^4\c^4}{2s_*^2\sqrt{1-s_{2\theta}^2X}}AX\right)\right]\nn\\
x^2&=&2s_\theta^2c_\theta^2X\frac{M_Z^2}{M_1^2(1-z^2)}
\eea
Now that we have expressed all in terms of
$e,\theta,M_Z,z,b_1,b_2,g_1$, we can rewrite the coupling
between the $Z$-boson and fermions (from Eq. (\ref{az})) as:
\be
a_Z^f=-\frac{e}{\s
\c}\left(1+\frac{\Delta\rho}{2}\right)\left(\mathbf{T_3}^f-s_{eff}^2\mathbf{Q}^f\right)
\ee
where
\bea
\!\!\!\!\!\!\!\!\!\!\!\!\!\!\!\!1+\frac{\Delta\rho}{2}
&=&\frac{1}{\sqrt{X}}\left(1-\frac{b}{2}\right)\left[1+
\frac{x^2}{s_*^2}\left(z_{Zb}-
\left(1-\frac{b}{2}\right)^2\frac{z_{Wb}}{X}+\frac{B}{2X}\right)\right]
\eea
and
\be
s_{eff}^2=\tilde s^2\frac{1}{1-\frac b 2}\left[1+\tilde x^2(\tilde c^2-\tilde
s^2-z_{Zb})\right]
\ee
Therefore
\be
\epsilon_1=\Delta\rho
=-2+\frac{2}{\sqrt{X}}\left(1-\frac{b}{2}\right)+\frac{2e^2}{s_*^2g_1^2\sqrt{X}}\left(1-\frac{b}{2}\right)
\left[z_{Zb}-\left(1-\frac{b}{2}\right)^2\frac{z_{Wb}}{X}+\frac{B}{2X}\right]
\label{drho}
\ee
with $z_{Z}$ and $z_{Zb}$ given in Eq.~(\ref{zz}) and (\ref{zzb}) with $\tilde s\to s^*$ and $\tilde c\to c^*$.
From the definition:
\be
s_{eff}^2=s^2_\theta (1+\Delta k)
\ee
 we get:
\be
\Delta k=-1+\frac{s_*^2}{\s^2}\frac{1}{1-\frac b
2}\left[1+\frac{x^2}{s_*^2}\left(\frac{\s^4\c^4}{\sqrt{1-s_{2\theta}^2X}}AX
+c_*^2-s_*^2-z_{Zb}\right)\right]
\label{dk}
\ee
Furthermore, from $M_W/M_Z$ we extract $\Delta r_W$:
\be
\frac{M_W^2}{M_Z^2}=\tilde c^2\left[1+\tilde x^2(z_Z-z_W)\right]=\c^2(1- \frac
{\s^2}{c_{2\theta}}\Delta r_W)
\ee
and
\be
\Delta r_W=\frac{\c^2-\s^2}{\s^2}\left\{
1-\frac{c_*^2}{\c^2}\left[1+x^2\left(\frac{z_Z}{s_*^2}-\frac{z_W}{s_*^2}
-\frac{\s^4\c^4}{c_*^2\sqrt{1-s_{2\theta}^2X}}AX\right)\right]\right\}
\label{drw}
\ee
Using Eqs.~(\ref{drho}),(\ref{dk}),(\ref{drw}) we derive $\varepsilon_{2,3}$ by using the relations in \cite{Altarelli:1991zd}:
\bea
\epsilon_2&=& \c^2 \Delta\rho+\frac {\s^2}{c_{2\theta}}\Delta r_W- 2 \s^2 \Delta k
\nn \\
\epsilon_3&=&\c^2\Delta\rho+ c_{2\theta}\Delta k
\eea


\hyphenation{Post-Script Sprin-ger}

\end{document}